\documentclass[journal]{IEEEtran}

\usepackage{cite}

\usepackage{amsmath}
\usepackage{mathrsfs}

\usepackage{algorithmic}

\usepackage{array}

\ifCLASSOPTIONcompsoc
\usepackage[caption=false,font=normalsize,labelfont=sf,textfont=sf]{subfig}
\else
\usepackage[caption=false,font=footnotesize]{subfig}
\fi

\usepackage{url}
\usepackage{graphicx,amsmath,amssymb,amsfonts}
\usepackage{algorithmic,algorithm}

\usepackage{setspace}

\usepackage{xcolor}

\newtheorem{remark}{\textbf{Remark}}
\newtheorem{theorem}{\textbf{Theorem}}

\newtheorem{lemma}{\textbf{Lemma}}

\newtheorem{corollary}{\textbf{Corollary}}

\newtheorem{proposition}{\textbf{Proposition}}

\renewcommand{\IEEEQED}{\IEEEQEDopen}

\makeatletter
\newcommand{\rmnum}[1]{\romannumeral #1}
\newcommand{\Rmnum}[1]{\expandafter\@slowromancap\romannumeral #1@}
\makeatother

\begin{document}
	% reference control
	\bstctlcite{ref:BSTcontrol}

	\title{Federated Learning in Multi-RIS Aided Systems}

	\author{Wanli~Ni,~Yuanwei~Liu,~Zhaohui~Yang,~Hui~Tian,~and~Xuemin~Shen
%		Wanli~Ni,~%\IEEEmembership{Student~Member,~IEEE,}
%		Yuanwei~Liu,~\IEEEmembership{Senior~Member,~IEEE,}
%		Zhaohui~Yang,~\IEEEmembership{Member,~IEEE,}
%		Hui~Tian,~\IEEEmembership{Senior~Member,~IEEE,}
%		and~Xuemin~Shen,~\IEEEmembership{Fellow,~IEEE}% <-this % stops a space
		\thanks{A short version of this work was presented at the IEEE GLOBECOM Workshop on Edge Learning over 5G Networks and Beyond, Taipei, Taiwan, Dec. 2020 \cite{Ni2012Intelligent}. }% <-this % stops a space
		\thanks{W. Ni and H. Tian are with the State Key Laboratory of Networking and Switching Technology, Beijing University of Posts and Telecommunications, Beijing, China (e-mail: charleswall@bupt.edu.cn; tianhui@bupt.edu.cn).}% <-this % stops a space
		\thanks{Y. Liu is with the School of Electronic Engineering and Computer Science, Queen Mary University of London, London, UK (e-mail: yuanwei.liu@qmul.ac.uk).}
		\thanks{Z. Yang is with the Centre for Telecommunications Research, Department of Engineering, King's College London, London, UK (e-mail: yang.zhaohui@kcl.ac.uk).}
		\thanks{X. Shen is with the Department of Electrical and Computer Engineering, University of Waterloo, Waterloo, Canada (e-mail: sshen@uwaterloo.ca).}% <-this % stops a space
	}
	
% make the title area
\maketitle

\begin{abstract}
	This paper investigates the problem of model aggregation in federated learning systems aided by multiple reconfigurable intelligent surfaces (RISs).
	The effective integration of computation and communication is achieved by over-the-air computation (AirComp), which can be regarded as one of uplink non-orthogonal multiple access (NOMA) schemes without individual information decoding.
	Since all local parameters are transmitted over shared wireless channels, the undesirable propagation error inevitably deteriorates the performance of global aggregation.
	The objective of this work is to
	\rmnum{1}) reduce the signal distortion of AirComp;
	\rmnum{2}) enhance the convergence rate of federated learning.
	Thus, the mean-square-error (MSE) and the device set are optimized by designing the transmit power, controlling the receive scalar, tuning the phase shifts, and selecting participants in the model uploading process.
	To address this challenging issue, the formulated mixed-integer non-linear problem (P0) is decomposed into a non-convex problem (P1) with continuous variables and a combinatorial problem (P2) with integer variables.
	In an effort to solve the MSE minimization problem (P1), the closed-form expressions for transceivers are first derived, then the multi-antenna cases are addressed by the semidefinite relaxation.
	Next, the problem of phase shifts design is tackled by invoking the penalty-based successive convex approximation method.
	In terms of the combinatorial optimization problem (P2), the difference-of-convex programming is adopted to optimize the device set for convergence acceleration, while satisfying the aggregation error demand.
	After that, an alternating optimization algorithm is proposed to find a suboptimal solution for the original non-convex problem (P0), where the corresponding convergence and complexity are analyzed.
	Finally, simulation results demonstrate that
	\rmnum{1}) the designed algorithm can converge faster and aggregate model more accurately compared to baselines;
	\rmnum{2}) the training loss and prediction accuracy of federated learning can be improved significantly with the aid of multiple RISs.
\end{abstract}


\section{Introduction}
%\IEEEPARstart{T}{his} demo file is intended to serve as a ``starter file''.
As one of the most promising frameworks of distributed machine learning, federated learning enables geo-distributed Internet-of-Thins (IoT) devices to collaboratively perform model training while keeping the raw data processed locally \cite{Shi2020Communication, Lyu2019Optimal}.
By doing so, federated learning has its unique advantages over centralized learning paradigms \cite{Liu2020When, Chen2021PNAS}.
Firstly, federated learning can effectively avoid the transmission of privacy-sensitive data over the wireless channels and is able to keep the collected data stored at different IoT devices, which is beneficial to preserve user privacy and data security \cite{Li2020SPM, Chen2021PNAS, Liu2020Privacy}.
Secondly, due to the fact that learning devices only need to communicate with the base station (BS) on the up-to-date model parameters \cite{Chen2021PNAS}, thus the communication overhead can be significantly reduced in a distributed learning fashion, which helps to overcome the drawback of excessive propagation delay caused by the potential network congestion \cite{Li2020SPM}.
Thirdly, exploiting the superposition property of multiple-access channel (MAC), over-the-air computation (AirComp) can be adopted to complete the local parameter communication and global model computation processes via concurrent transmission \cite{Nazer2007Computation}.
Broadly speaking, AirComp without individual information decoding can be regarded as one of uplink non-orthogonal multiple access (NOMA) techniques \cite{Liu2020Privacy, Zhu2020Broadband}, and thus both the completion time and spectrum efficiency of the over-the-air federated learning (AirFL) system can be improved in comparison with the conventional orthogonal multiple access-based schemes \cite{Chen2021PNAS, Chen2021Convergence}.
Lastly, compared with the conventional cloud learning, federated learning is inherently conducive to offloading compute-intensive tasks from the central server to the edge devices \cite{Zhu2020Toward, Chen2020Wireless}, which can speed up the processing of real-time data by making full use of the dispersed computation resources at the network edge.
However, owing to the resource-limited IoT devices and the non-uniform fading channels \cite{ChenQW18WCL}, problems such as the signal distortion and aggregation error will seriously degrade the convergence rate and prediction accuracy of the federated learning system.
Therefore, it is important to design innovative, spectrum-efficient, and communication-efficient solutions for the federated learning over IoT networks.

By installing a large number of passive reflecting elements on the programmable surfaces, reconfigurable intelligent surfaces (RISs, also known as intelligent reflecting surfaces, relay 2.0, etc.) have been recognized as a novel technology to smartly reconfigure the complex propagation environment of radio signals \cite{Liu2021Reconfigurable}.
Specifically, through judiciously controlling the amplitude and phase shift of each reflecting element in real time, RISs are able to proactively modify the wireless channels between the BS and devices, and there is no need for complicated interference management even if multiple RISs are considered \cite{Yang2021Energy}.
Moreover, although traditional active relays that support multiple-input multiple-output (MIMO) or millimeter-wave communication can achieve similar effects, RISs have better performance in terms of hardware cost and energy consumption \cite{Wu2019IRS}.
Thereby, the software-controlled RISs provide a new paradigm for realizing a smart and programmable wireless environment and then further improving the performance of existing networks.
Nevertheless, the ever-increasing complexity of wireless networks composed of a set of heterogeneous facilities makes effective modeling and networking difficult if not impossible.
Hence, the efficient deployment of RIS-aided intelligent IoT networks faces challenges from system characterization to performance optimization \cite{Yang2021Energy, Wu2019IRS, Ni2021Integrating}.

Sparked by the aforementioned benefits and issues of AirFL and RISs, it is imperative and valuable to integrate them together to reduce the propagation error of distributed learning and accelerate the convergence rate of global aggregation, due to the following profits and reasons:
\begin{itemize}
	\item First of all, as an uplink NOMA scheme, the performance of AirFL can be improved by flexibly designing the phase shifts of RISs to exploit the superposition property of MAC channels.
	Through aligning multiple signals via RISs, MAC channels can be deemed as a virtual computer capable of matching the desired aggregation function of AirFL.
	The efficient combination of communication and computation also helps to boost the spectrum utilization of resource-constrained IoT networks.
	\item Then, the signal distortion of AirComp can be further reduced by deploying multiple RISs to merge reflected signals dexterously, so that	the model parameters from local users can be aggregated more accurately.
	Another benefit of RISs is to provide available links for cell-edge users blocked by obstacles, thereby enhancing the coverage and connectivity of federated learning.
	After all, more learning participants with reliable channels can speed up the convergence rate of global aggregation.
	\item
	Last but not least, compared with conventional active relays, RISs usually do not require dedicated energy supplies for operation.
	Therefore, RISs can be easily integrated into existing wireless networks without changing any standard or hardware, so that the energy efficiency of conventional IoT networks can be enhanced significantly without increasing extra huge operating expenses.
\end{itemize}

\subsection{Related Works}
Recently, both federated learning and RISs have attracted remarkable attention and have been implemented separately in various application scenarios.
So far, the majority of previous works such as \cite{Zhu2020Broadband, Xu2021Learning, Wang2020Federated, Yang2020Federated, Amiri2020Federated, Yang2021Energy, Chen2020A, Chen2018Over, Liu2020Over, Yang2020TWC} have studied the implementation of federated learning over wireless networks.
%
% FL: latency, Wireless Fading Channels
Specifically, by implementing distributed stochastic gradient descent for parameter updating, Amiri \textit{et al.} \cite{Amiri2020Federated} proposed digital and analog communication schemes for federated learning over a shared bandwidth-limited fading MAC.
%
% FL: latency, broadband analog aggregation
In order to shorten the communication latency, Zhu \textit{et al.} \cite{Zhu2020Broadband} proposed a broadband analog aggregation scheme for federated learning, which outperformed the conventional orthogonal access.
%
% FL: latency, energy efficiency
Furthermore, Yang \textit{et al.} \cite{Yang2021Energy} investigated the resource allocation problem of joint federated learning and wireless communication to strike a trade-off between completion time and energy consumption for edge devices.
%
% FL: error, resource allocation
For the purpose of minimizing the training error of federated learning, Chen \textit{et al.} \cite{Chen2020A} derived a closed-form expression for the expected convergence rate.
%
% FL: error
Taking both the intra-node interference and the non-uniform fading into account, the authors in \cite{Chen2018Over} analyzed the aggregation performance of AirComp and derived the closed-form expression of the mean-square-error (MSE) outage, then receive antenna selection was adopted to avoid massive channel state information (CSI) gathering in the MIMO networks.
%
% FL: error, AirComp MSE
Exploiting the superposition property of MAC and the functional decomposition, Liu \textit{et al.} \cite{Liu2020Over} focused on the MSE minimization problem of AirComp by designing the transceiver policy under the power constraint, where the closed-form expressions for computation-optimal strategy were derived.
%
% FL: error + latency accelerating model aggregation
With the aim of accelerating model aggregation and reducing test error, the authors in \cite{Yang2020TWC} jointly optimized the device selection and receive vector to improve convergence rate and prediction accuracy of federated learning.
%
% FL: error + latency: Multi-Objective Evolutionary Federated Learning
%From the perspective of modifying the neural network structure in federated learning, the authors of \cite{Zhu2020Multi} adopted a multi-objective evolutionary algorithm to prune the connections between neurons for communication cost reduction and learning accuracy maximization.
%
% FL: network lifetime
%To prolong the network lifetime of densely deployed Internet of Things (IoT) networks, Basaran \textit{et al.} \cite{Basaran2020Energy} presented a novel estimator to provide a MSE improvement with reducing energy consumption.
%
% FL + WPT
%Integrating the wireless powered transfer and AirComp techniques, Li \textit{et al.} \cite{Li2019Wirelessly} jointly optimized the wireless power control and aggregation beamforming to minimize aggregation error.

%
Meanwhile, several basic challenges with respect to (w.r.t.) RIS-aided communications have been solved in a number of prior works such as \cite{Wu2019IRS, Huang2019EE, Hou2020RIS, Xie2020Max, Huang2020Reconfigurable, Liu2020RIS, Ni2021Resource, Yang2021Energy}.
With the objective to maximize the energy efficiency in the RIS-incorporated cellular systems, Huang \textit{et al.} \cite{Huang2019EE} jointly optimized the active and passive beamforming in a downlink multi-user communication network.
% minimize the transmit power 
%With the objective to minimize the transmit power at the access point, Wu \textit{et al.} \cite{Wu2020Beamforming} alternately designed the active and passive beamforming in the single-user and multi-user cases with single IRS, where the performance loss of discrete phase shifts was analyzed as well.
%
% RIS: MIMO-NOMA
By deploying RISs to eliminate the inter-cluster interference in MIMO-NOMA networks for performance enhancement, Hou \textit{et al.} \cite{Hou2020RIS} obtained the minimal required number of RISs for the signal cancellation demand.
%
% RIS: multi-RIS, weighted sum-rate
Considering the user fairness in RIS-aided systems, the max-min problem was optimized in \cite{Xie2020Max} by designing the transmit power and phase shifts in an iterative manner.
%
% RIS: DRL, sum-rate maximization
Unlike the alternating optimization, to solve the high-dimension problem of the sum-rate maximization in RIS-assisted MIMO systems, Huang \textit{et al.} \cite{Huang2020Reconfigurable} leveraged the deep reinforcement learning (DRL) to obtain the joint design of the transmit beamforming and the reflection matrix.
%
% RIS: DRL, energy efficiency
Similarly, using DRL approaches, an agent for determining the position and phase shifts of RIS was trained in \cite{Liu2020RIS} to maximize the long-term energy efficiency of NOMA networks by learning the optimal control strategy in a trial-and-error manner.
%
% RIS: throughput
Additionally, considering the problem of resource allocation in the RIS-aided NOMA networks, our previous work in \cite{Ni2021Resource} jointly optimized the phase shifts, transmit power, user pairing and subchannel assignment to maximize system throughput.
%
% RIS: energy efficiency
With the aid of multiple RISs, the work in \cite{Yang2021Energy} maximized the energy efficiency by dynamically controlling the on-off states of RISs and iteratively optimizing their corresponding phase shifts.

%
%% theoretical analysis
%By considering the perfect and imperfect decoding of the RIS-aided network, the authors of \cite{Yue2020Analysis} and \cite{Hou2019IRS} gave the closed-form expressions for the outage probability and ergodic rate.
%%
%% theoretical analysis
%Due to the hardware limitations of RIS in practice, the impacts of finite-resolution amplitude and phase shifts on outage probability and achievable data rate were analyzed in \cite{Ding2020IRS} and \cite{Zhang2020IRS}, respectively.
%%
%% throughput
%By considering the ergodic and delay-limited capacity for RIS-aided networks, Mu \textit{et al.} \cite{Mu2020Capacity} jointly optimized the phase shifts and resource allocation to maximize the average sum rate of all users by invoking Lagrange duality method.
%%
%% energy efficiency
%Compared to common multi-antenna amplify-and-forward (AF) relays, Huang \textit{et al.} \cite{Huang2019EE} demonstrated that RIS-aided networks can enjoy a higher energy efficiency.
%%
%% energy efficiency
%Base on the deep reinforcement learning approaches, an agent for determining the position and phase shifts of RIS was trained in \cite{Liu2020RIS} to maximize the long-term energy efficiency with data rates requirements by learning the optimal control strategy in a trial-and-error manner.

%\newpage
\subsection{Challenges and Contributions}
Since artificial intelligence (AI) plays a defining role in the design of future 6G networks, RIS-aided federated learning can be deemed as an attractive candidate actualizing the efficient integration of distributed learning and wireless IoT networks, which caters to the needs of next-generation wireless networks supporting massive connectivities of AI.
However, there is still a paucity of research contributions on investigating AirFL systems aided by intelligent surfaces, especially for the multi-objective problem w.r.t. aggregation distortion and device selection in AirFL, thereby motivating this work.
Before exploiting specific techniques to improve federated learning performance over non-ideal wireless IoT networks, we first summarize potential challenges as follows:
\begin{itemize}
	\item
	So far, it is a highly challenging issue to minimize MSE by jointly designing the transmit power, receive scalar, and reflection coefficients in a communication-efficient manner, while guaranteeing the global aggregation error requirements within the available power budget.
	\item Moreover, one can know that the combinational optimization w.r.t. the device selection problem is non-deterministic polynomial-time (NP) hard.
	The complexity of exhaustive search is exponential, so it is non-trivial to obtain an optimal solution in polynomial time.
\end{itemize}

%\subsection{Contributions}
In order to reduce propagation error while speeding up convergence rate, we jointly optimize model synchronization and device selection problems in AirFL systems using multiple RISs.
Notably, RISs play a vital role in turning the wireless channels into a functional computer to better match the desired weighted sum feature of federated learning.
More precisely, multiple geo-distributed RISs are deployed to enhance the parameter aggregation from IoT devices to the BS in an efficient manner.
Due to the non-convexity of the objective function and constraints, the formulated problem is difficult to solve optimally.
For the purpose of expanding our contributions, both single-antenna and multi-antenna cases are considered, in which the closed-form solutions and iterative algorithms are developed to obtain solutions with high performance.
To the best of our knowledge, RIS-aided AirFL is still at its nascent stage and many open issues remain to be addressed such as the joint design of transmit-reflect-and-receive in multi-RIS aided IoT networks.
The main contributions of this work can be summarized as follows:
\begin{enumerate}
	\item 
	We propose a framework of resource allocation and device selection in the AirFL system for model aggregation with the aid of multiple RISs.
	Accordingly, we formulate a bi-criterion problem for aggregation accuracy enhancement and convergence rate improvement by jointly optimizing the transmit power, receive scalar, phase shifts, and device set, subject to power constraints of devices and unit-modulus constraints of RISs as well as the aggregation error requirement.
	Meanwhile, we analyze that the original problem is a mixed-integer non-linear programming (MINLP) problem, which is NP-hard and is non-trivial to solve directly.
	\item
	In order to tackle the non-convex MSE minimization problem with continuous variables, we first derive the closed-form expressions for transceiver designs. Next, we adopt methods such as semidefinite relaxation (SDR) and successive convex approximation (SCA) to transform the non-convex subproblems into convex ones, and then solve them in polynomial time complexity.
	Afterwards, we invoke difference-of-convex (DC) programming to handle the cardinality maximization problem with combinatorial features.
	Subsequently, we propose an alternating optimization algorithm to solve the original MINLP problem and analyze the corresponding convergence and complexity.
	\item
	We conducts comprehensive simulations on the synthetic and real datasets to validate the effectiveness of our designed algorithms.
	Numerical results show that the proposed communication-efficient solutions for RIS-aided AirFL systems outperform benchmarks, such as single-RIS and random-phase schemes.
	Specifically, our algorithms can achieve better convergence rate and lower learning error in the experiments of implementing AirFL for linear regression and image classification.
	Meanwhile, we verify that the deployment of RISs is beneficial to alleviate propagation error and reduce signal distortion of AirFL over shared wireless channels.
%	The designed algorithms are also capable of reducing energy consumption and prolonging the network lifetime.
\end{enumerate}

\subsection{Organization and Notation}
The rest of this paper is organized as follows.
First, the system model of multi-RIS aided AirFL is given in Section \ref{system}.
Then, a bi-criterion optimization problem is formulated in Section \ref{problem_formulation} where the problem decomposition is conducted as well.
Next, an alternating optimization algorithm is proposed in Section \ref{alternating}. 
The corresponding convergence and complexity are analyzed in Section \ref{convergence}.
Finally, extensive numerical simulations are presented in Section \ref{simulation}, which is followed by conclusion in Section \ref{conclusion}.

The key notations of this paper are summarized as follows.
Scalars are denoted by italic letters, vectors and matrices are denoted by bold-face lower-case and uppercase letters, respectively.
The space of $m \times n$ complex-valued matrices is denoted by $\mathbb{C}^{m \times n}$.
The distribution of a complex Gaussian random vector with mean vector $\mu$ and covariance matrix $\sigma^2$ is denoted by $\mathcal{CN}(\mu,\sigma^2)$, and $\sim$ stands for "distributed as".
For a complex number $x$, the amplitude is denoted by $|x|$.
Meanwhile, ${\rm Re}(x)$ and ${\rm Im}(x)$ denote the real and imaginary parts of $x$, respectively.
For a complex vector $\boldsymbol{x}$, ${\rm diag} ( \boldsymbol{x} )$ denotes a diagonal matrix with each diagonal element being the corresponding element in $\boldsymbol{x}$.

\section{System Model} \label{system}

\begin{figure*} [t!]
	\centering
	\includegraphics[width=5.2 in]{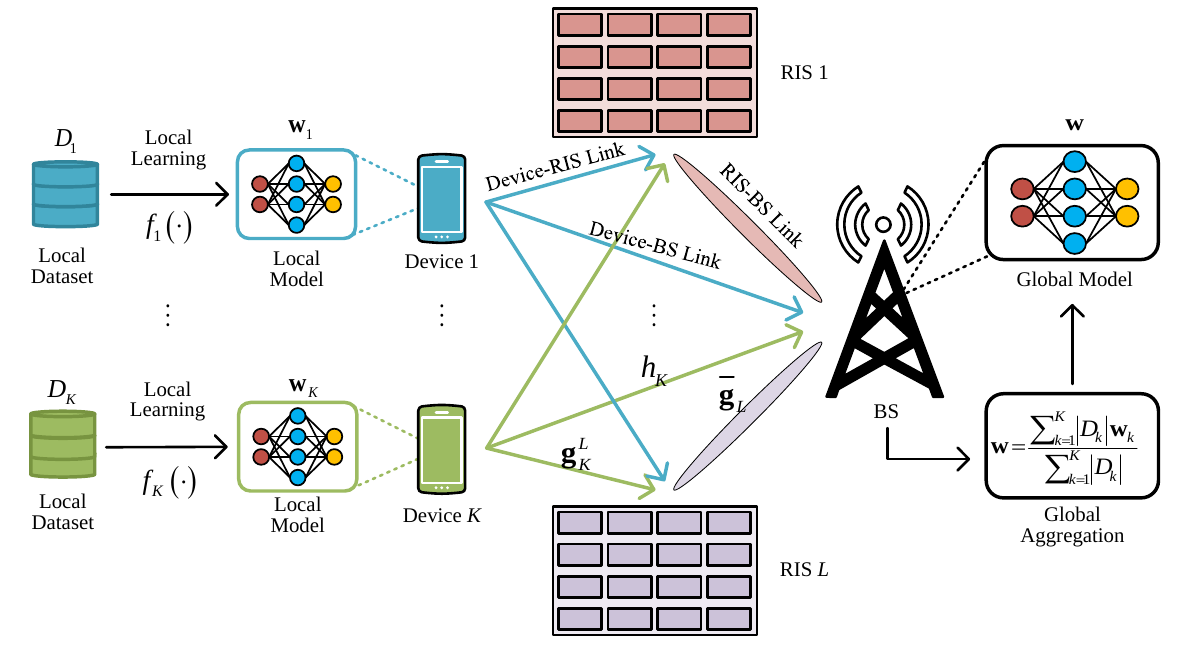}
	\caption{An illustration of over-the-air federated learning in multi-RIS aided IoT networks.}
	\label{system_model}
\end{figure*}

As illustrated in Fig. \ref{system_model}, we consider a RIS-aided AirFL system consisting of one BS, $N$ intelligent IoT devices, and $L$ RISs.
Assume that both the BS and devices are equipped with a single antenna, and each RIS comprises $M$ reflecting elements.
Instead of aggregating all local parameters, the number of devices selected to participate in the model uploading process is $K$ out of $N$ $( 1 \le K \le N )$.
The sets of selected devices and RISs are indexed by $\mathcal{K} = \{ 1,2, \ldots, K\}$ and $\mathcal{L} = \{ 1,2, \ldots, L\}$, respectively.
Let $\mathcal{D} = \{\mathcal{D}_{1},\mathcal{D}_{2},\ldots,\mathcal{D}_{K}\}$ denote the dataset collected by all selected devices, where $\mathcal{D}_{k}$ is the raw data recorded by the $k$-th device.
The diagonal matrix of the $\ell$-th RIS is denoted by $\mathbf{\Theta}_{\ell} = \text{diag} ( e^{j \theta_{\ell}^{1}}, e^{j \theta_{\ell}^{2}}, \ldots, e^{j \theta_{\ell}^{M}} )$, where $\theta_{\ell}^{m} \in [0, 2 \pi]$ denotes the phase shift of the $m$-th reflecting element on the $\ell$-th RIS\footnote{In practice, each RIS can communicate with the BS via a separate link connected by a programmable controller that is capable of smartly adjusting the phase shifts of all reflecting elements in real time \cite{Yang2021Energy, Wu2019IRS, Yang2020Federated}.}.

\begin{figure*} [t!]
	\centering
	\includegraphics[width=6 in]{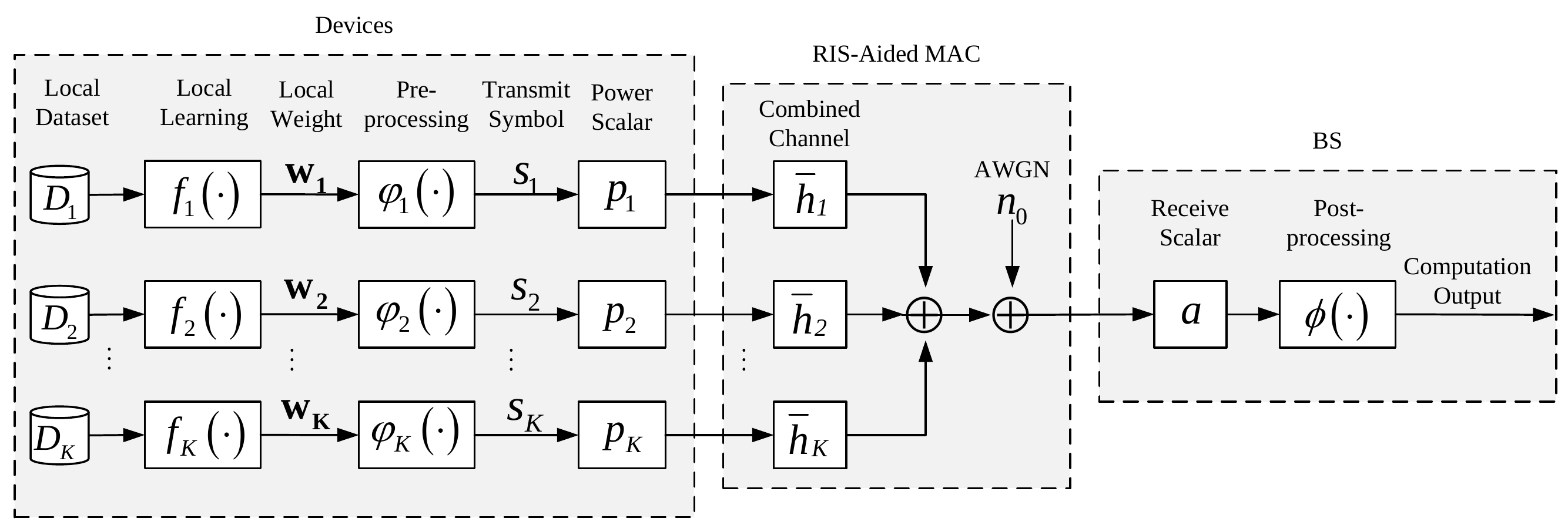}
	\caption{Block diagram of federated learning via over-the-air computation.}
	\label{block_diagram_AirComp}
\end{figure*}

The block diagram of AirFL is illustrated in Fig. \ref{block_diagram_AirComp}, which can be deemed as a function-centric uplink NOMA technique that does not need to decode users' information one by one.
All devices transmit their up-to-date local models $\{\mathbf{w}_{k} \mid \forall k \in \mathcal{K}\}$ simultaneously over the same time-frequency resource\footnote{Unlike the conventional communicate-then-compute schemes, AirComp is capable of harnessing the channel interference to integrate communication and computation into one concurrent transmission \cite{Chen2018Over, Liu2020Over, Yang2020TWC}. Moreover, all selected devices is assumed to be well synchronized so that the signals are perfectly aligned \cite{Jiang2019Over, Wang2019Adaptive, Zhu2019AirComp}, the analysis of asynchronous transmission is beyond the scope of this paper.},
then the target function computed at the BS can be written as \cite{Liu2020Over}
\begin{equation}
\psi \left( \mathbf{w}_{1}, \mathbf{w}_{2},\ldots, \mathbf{w}_{K} \right) = \phi \left( \sum \nolimits_{k=1}^{K} \varphi_k \left( \mathbf{w}_{k} \right) \right),
\end{equation}
where $\mathbf{w}_{k} = f_{k}(\mathcal{D}_{k})$ is the updated local model at the $k$-th device, $\varphi_k (\cdot)$ and $\phi (\cdot)$ denote the pre-processing function and the post-processing function, respectively. 
Before the BS computes the target function $\psi (\cdot)$, it needs to collect the target-function variable $s$, defining as
\begin{equation} \label{target_variable}
s = \sum \nolimits_{k=1}^{K} s_k \quad \text{and} \quad s_k = \varphi_{k} \left( \mathbf{w}_{k} \right),
\end{equation}
where $s_k \in \mathbb{C}$ is the transmit symbol after pre-processing at the $k$-th device.

Let $h_k \in \mathbb{C}$, $\mathbf{g}_{k}^{\ell} \in \mathbb{C}^{M \times 1}$, and $\mathbf{\bar{g}}_{\ell} \in \mathbb{C}^{1 \times M}$ denote the channel responses from the $k$-th device to the BS, from the $k$-th device to the $\ell$-th RIS, and from the $\ell$-th RIS to the BS, respectively.
Using the AirComp technique, the received superposition signal at the BS can be given by
\begin{equation}
{y} = \sum_{k=1}^{K} \left( {h}_{k} + \sum_{\ell=1}^{L} \mathbf{\bar{g}}_{\ell} \mathbf{\Theta}_{\ell} \mathbf{g}_{k}^{\ell} \right) {p_{k}} s_{k} + n_{0},
\end{equation}
where
$p_{k} \in \mathbb{C}$ is the transmit power scalar at the $k$-th device, 
$n_{0} \sim \mathcal{CN}(0, \sigma^2 )$ is the additive white Gaussian noise (AWGN), and $\sigma^2$ is the noise power.

The transmit symbols are assumed to be independent and are normalized with unit variance, i.e.,
$\mathbb{E}(s_k^{ \rm H}s_{k'})=0, \forall k \ne k'$ and $\mathbb{E}(|s_k|^2)=1$.
Then, the transmit power constraint at the $k$-th device can be given by 
\begin{equation}\label{power_constraint}
\mathbb{E}(| {p_{k}} s_k |^2) = | p_k |^2 \le P_0, \ \forall k \in \mathcal{K},
\end{equation}
where $P_{0} > 0$ is the available maximum transmit power of each device.

By employing a receive scalar ${a} \in \mathbb{C}$ to the received signal $y$, the estimation at the BS is thus given by
\begin{equation} \label{estimation}
\hat{s} = \frac{1}{\sqrt{\eta}} {a} {y}
 		= \frac{a}{\sqrt{\eta}} \sum_{k=1}^{K} \bar{h}_{k} {p_{k}} s_{k}
 		+ \frac{{a}}{\sqrt{\eta}} n_{0},
\end{equation}
where
$\bar{h}_{k} = {h}_{k} + \sum_{\ell=1}^{L} \mathbf{\bar{g}}_{\ell} \mathbf{\Theta}_{\ell} \mathbf{g}_{k}^{\ell}$ is the combined channel, and $\eta > 0$ is a normalizing factor.

Comparing the target-function variable $s$ in (\ref{target_variable}) with the observed one $\hat{s}$ in (\ref{estimation}), the corresponding error can be calculated by $e = \hat{s} - s$.
In this paper, the MSE is adopted as the performance metric of AirFL, which is different from the rate-centric NOMA transmission in the literature \cite{Hou2020RIS,Xie2020Max,Huang2020Reconfigurable,Liu2020RIS,Ni2021Resource} that aim to maximize the system throughput or the individual rate.
Specifically, to quantify the performance of AirComp in global aggregation, the distortion of $\hat{s}$ w.r.t. $s$ is estimated by the MSE defined as
\begin{equation} \label{MSE_s}
\begin{aligned}
\text{MSE} (\hat{s}, s)
\triangleq \mathbb{E} ( |\hat{s} - s|^2)
= \sum_{k=1}^{K} \left| \frac{1}{\sqrt{\eta}} {a} \bar{h}_{k} {p_{k}} - 1 \right| ^2 + \frac{\sigma^2 | {a} |^2 }{\eta}.
\end{aligned}
\end{equation}

Note that the first-order Taylor approximation of the computed target function $\hat{\psi} = \phi(\hat{s})$ at $s$ can be rewritten by
\begin{equation}
\hat{\psi} \approx \phi(s) + \phi'(s) (\hat{s} - s).
\end{equation}
Then, with given $\phi'(s)$, the equivalent transformation between the MES of $\psi$ and the MSE of $s$ can be expressed as
\begin{equation}
\text{MSE}(\hat{\psi}, \psi) \approx |\phi'(s)|^2 \ \text{MSE}(\hat{s}, s),
\end{equation}
which implies that a minimum MES of $\psi$ also leads to a minimum MSE of $s$.
At this point, it can be concluded that the minimization of (\ref{MSE_s}) is a reasonable surrogate of the minimum $\text{MSE}(\hat{\psi}, \psi)$.
Thus, $\text{MSE} (\hat{s}, s)$ is regarded as one of the performance metrics in the rest of this paper.

\section{Problem Formulation and Decomposition} \label{problem_formulation}
Given the considered system model of RIS-aided federated learning, both the aggregation error and convergence rate depend on resource allocation and device selection schemes.
Therefore, we shall investigate the optimization of transmit power, receive scalar, phase shifts, and learning participants to minimize MSE for prediction accuracy improvement, while selecting as more devices as possible for convergence accelerating.
%
%To this end, the objective of this work is to minimize the weighted sum of the objectives in terms of MSE and cardinality, thus the bi-criterion optimization problem can be formulated as
To this end, the bi-criterion optimization problem can be formulated as
\begin{subequations}\label{bi_criterion}
	\begin{eqnarray}
	(\mathcal{P}0): 
	\label{bi_criterion_objective}
	&\min \limits_{ \boldsymbol{p}, a, \boldsymbol{\theta}, \mathcal{K} } & \text{MSE} (\hat{s}, s) - \gamma \left| \mathcal{K} \right| \\
	\label{bi_criterion_power_constraint}
	&{\rm s.t.}&| p_k |^2 \le P_0, \ \forall k \in \mathcal{K}, \\
	\label{bi_criterion_phase_constraint}
	&{}&0 \le \theta_{\ell}^{m} \le 2 \pi, \ \forall \ell, m, \\ % \theta_{\ell}^{m} \in [0, 2\pi]
	\label{bi_criterion_MSE}
	&{}&\text{MSE} (\hat{s}, s) \le \varepsilon_0, \\
	\label{bi_criterion_participant}
	&{}&1 \le \left| \mathcal{K} \right| \le N,
	\end{eqnarray}
\end{subequations}
where
$\boldsymbol{p} = [p_1, p_2, \ldots, p_K]^{\rm T}$ is the transmit power vector,
$\boldsymbol{\theta} = \left[ \theta_{1}^{1}, \theta_{1}^{2}, \ldots, \theta_{1}^{M}, \theta_{2}^{1}, \theta_{2}^{2}, \ldots, \theta_{L}^{M} \right] ^{\rm T}$ is the phase shifts vector,
$\varepsilon_0 > 0$ is the aggregation error requirement,
$|\mathcal{K}|=K$ is the cardinality of set $\mathcal{K}$,
and $\gamma > 0$ is a problem parameter to achieve a trade-off between the aggregation error and the convergence rate.
%\footnote{According to the results obtained in \cite{Yang2020TWC}, it can be known that selecting more devices to participate the processes of parameter updating and model uploading is able to improve the convergence rate of federated learning. This is because selecting more devices means that more data samples can be used for federated learning, which can improve the learning performance.}
By adjusting the parameter $\gamma$, the optimal trade-off curve between $\text{MSE} (\hat{s}, s)$ and $|\mathcal{K}|$ can be swept out.
Additionally, the transmit power constraints are provided in (\ref{bi_criterion_power_constraint}).
The phase shift constraints are given in (\ref{bi_criterion_phase_constraint}).
The MSE tolerance of global aggregation is  presented in (\ref{bi_criterion_MSE}).
The number of learning participants is  limited in (\ref{bi_criterion_participant}).
%Before solving this intractable problem, the analysis of problem (\ref{bi_criterion}) is represented in the following.

The bi-criterion problem (\ref{bi_criterion}) is a MINLP problem due to the coupling of continuous variables and discrete variables in both objective function and constraints.
More specifically, the original problem (\ref{bi_criterion}) is still intractable even for the case without RISs, i.e., $L = 0$, due to the non-convex objective function and the combinatorial features of device selection.
One can know that it is highly intractable to directly find the global optimal solution of the NP-hard problem (\ref{bi_criterion}).
Upon this, in order to address this MINLP problem ($\mathcal{P}0$) effectively, we propose to decouple it into two subproblems ($\mathcal{P}1$) and ($\mathcal{P}2$).
Specifically, if the set of selected device $\mathcal{K}$ is fixed, problem (\ref{bi_criterion}) becomes subproblem (\ref{MSE_minimization}) of MSE minimization.
If the transmit power vector $\boldsymbol{p}$, the receive scalar $a$ and the phase shifts $\boldsymbol{\theta}$ are fixed, problem (\ref{bi_criterion}) becomes subproblem (\ref{device_selection}) of combinatorial optimization, i.e.,
\begin{enumerate}
	\item \emph{MSE minimization}: Given the set of device selection, the first objective is to minimize MSE by dynamically controlling the phase shifts of each RIS and optimizing the transmit power of each selected device as well as the receive scalar at the BS, subject to power constraints for devices and unit-modulus constraints for RISs.
	As a result, the corresponding MSE minimization subproblem is given by
	\begin{subequations}\label{MSE_minimization}
		\begin{eqnarray}
		(\mathcal{P}1): 
		\label{MSE_minimization_objective}
		&\min \limits_{ \boldsymbol{p}, a, \boldsymbol{\theta} } & \text{MSE} (\hat{s}, s) \\
		&{\rm s.t.}& {\rm (\ref{bi_criterion_power_constraint}) \ and \  (\ref{bi_criterion_phase_constraint})}.
		\end{eqnarray}
	\end{subequations}
	\item \emph{Combinatorial optimization}: Given the transmit power, receive scalar and phase shifts, the second objective is to minimize the aggregation error and maximize the number of selected devices at the same time by solving the following combinatorial optimization subproblem, subject to the MSE requirement for global aggregation and the cardinality constraint for participant number, which can be formulated as
	\begin{subequations}\label{device_selection}
		\begin{eqnarray}
		(\mathcal{P}2): 
		\label{device_selection_objective}
		&\min \limits_{ \mathcal{K} } & \text{MSE} (\hat{s}, s) - \gamma \left| \mathcal{K} \right| \\
		\label{MSE_requirement}
		&{\rm s.t.}& {\rm (\ref{bi_criterion_MSE}) \ and \ (\ref{bi_criterion_participant})}.
		\end{eqnarray}
		\end{subequations}
\end{enumerate}

\begin{figure*}
	\centering
	\subfloat[problem decomposition]{
		\label{roadmap} 
		\includegraphics[width=0.56\linewidth]{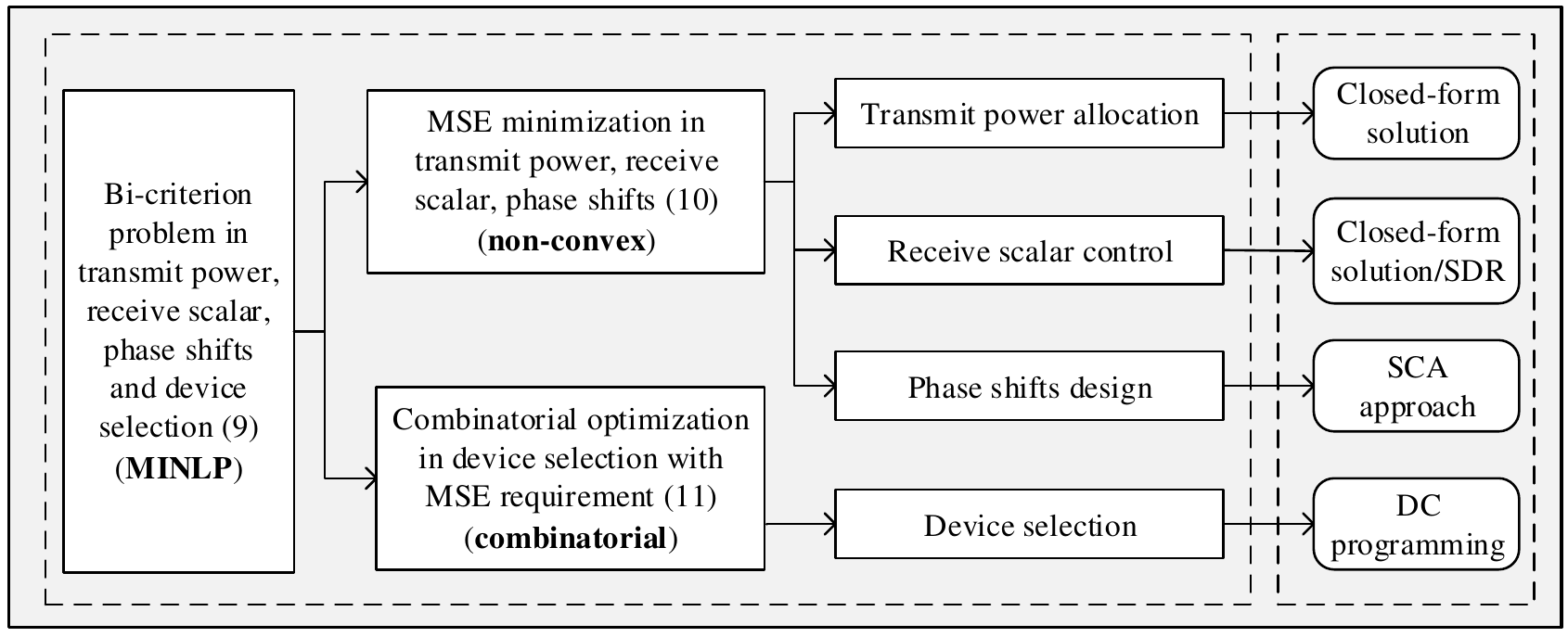}}
	\hspace{0 mm} 
	\subfloat[alternating optimization]{
		\label{flowchart}
		\includegraphics[width=0.34\linewidth]{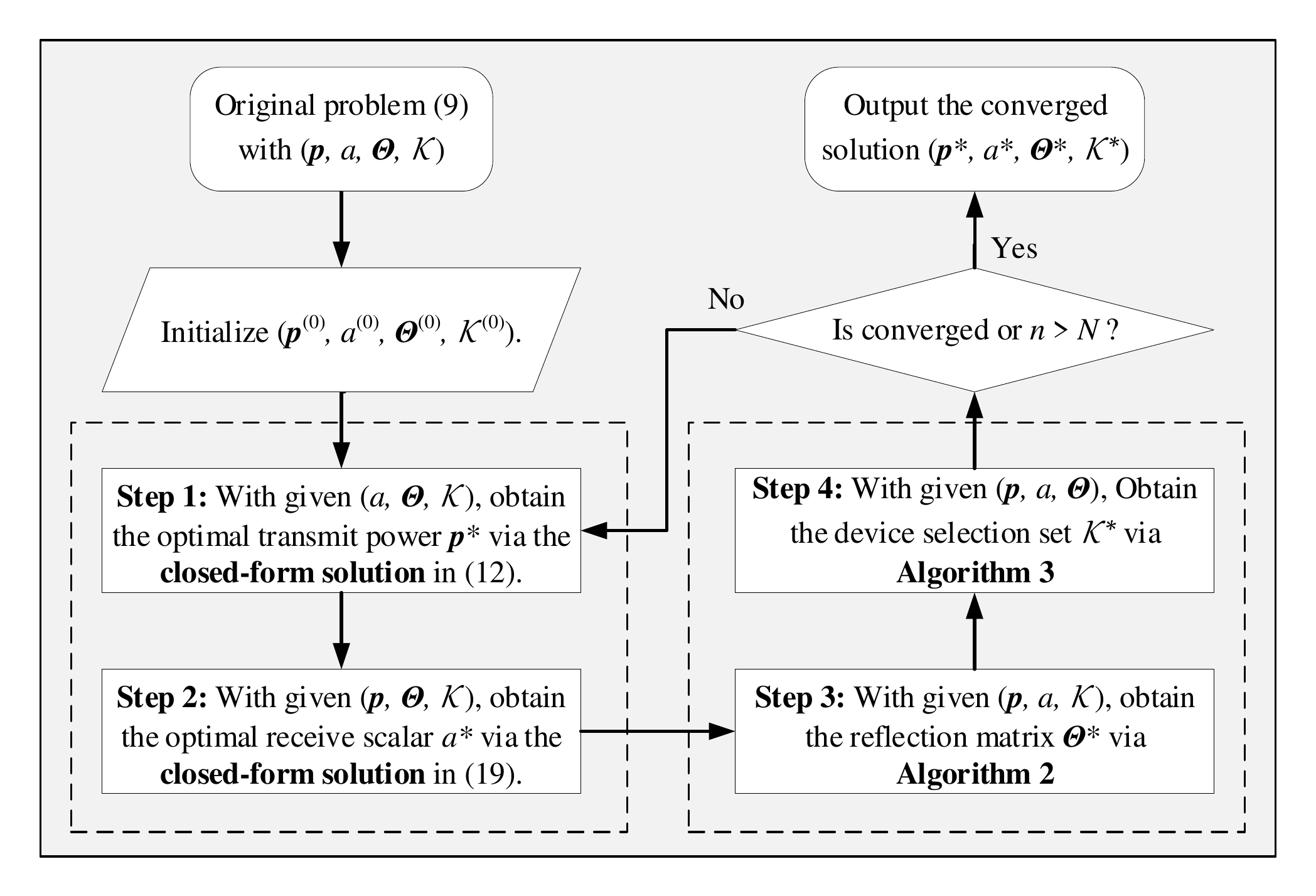}}
	\caption{Roadmap and flowchat: (a) an overview of the problem decomposition and proposed methods to subproblems; (b) a flowchart of the proposed alternating optimization algorithm.}
	\label{roadmap_and_flowchart} 
\end{figure*}

The MSE minimization problem (\ref{MSE_minimization}) is non-linear and non-convex even for the single-device case with $K=1$, due to the close coupling of $\boldsymbol{p}$, $a$ and $\boldsymbol{\theta}$ in $\text{MSE} (\hat{s}, s)$.
Not to mention that problem (\ref{MSE_minimization}) is still non-convex even when we only check the feasibility of phase shifts design.
Moreover, the combinatorial optimization problem (\ref{device_selection}) with multiple constraints is NP-hard and is non-trivial to obtain a high-performance solution as well.
Furthermore, problem (\ref{MSE_minimization}) is still a challenging issue even when some simplified cases are considered, while it is also almost impossible to obtain a closed-form solution to problem (\ref{device_selection}).
Fortunately, some common relaxation approaches can be adopted to transform the non-convex subproblems into convex ones, which can be solved separately and alternatively over iterations.
This motivates us to decompose the original problem into multiple subproblems. 
Thus, the alternating optimization method is invoked as an intuitive approach to solve the non-linear and non-convex problem (\ref{bi_criterion}) in an efficient manner, i.e., fix one and optimize the other, then repeat this in turn until the termination condition is satisfied.

However, due to the rapidly-varying CSI, it is impractical and not cost-effective for the resource-scarce devices to acquire global CSI when they allocate transmit power for model uploading.
To reduce the high signaling overhead of CSI feedback, it is of significance to develop a communication-efficient scheme for distributed power allocation.
Furthermore, one straightforward approach to find the optimal set of participating devices is the exhaustive search, but it inevitably results in an unacceptable computational complexity, i.e., $\mathcal{O} (2^{N})$.
As a result, to avoid the exponential complexity, it is imperative and desirable to design computation-efficient algorithms with polynomial time complexity.

%\begin{figure*}
%	\centering
%	\subfloat[problem decomposition]{
%		\label{roadmap} 
%		\includegraphics[width=0.56\linewidth]{roadmap.pdf}}
%	\hspace{0 mm} 
%	\subfloat[alternating optimization]{
%		\label{flowchart}
%		\includegraphics[width=0.34\linewidth]{flowchart.pdf}}
%	\caption{Roadmap and flowchat: (a) an overview of the problem decomposition and proposed methods to subproblems; (b) a flowchart of the proposed alternating optimization algorithm.}
%	\label{roadmap_and_flowchart} 
%\end{figure*}

%
To illustrate the problem decomposition, we provide a tree diagram in Fig. \ref{roadmap} to clearly delineate the connections between the key reformulated subproblems and the corresponding solutions.
Specifically, when the set of selected device is fixed, the non-convex MSE minimization subproblem (\ref{MSE_minimization}) is solved by the derived closed-form solutions and the developed SCA-based algorithm.
When the transceiver and reflection coefficients are fixed, finding the optimal solution is still non-trivial due to the combinatorial property of subproblem (\ref{device_selection}) in terms of the device selection.
Inspired by the DC representation method described in \cite{Gotoh2018DC}, a natural way to address it is invoking the DC programming.
Additionally, we provide a flowchart in Fig. \ref{flowchart} to draw the steps of the proposed alternating optimization algorithm for solving resource allocation and device selection problems in single-antenna cases.
The flowchart for multi-antenna cases can be obtained similarly.
Our specific solutions to subproblems in terms of transmit power, receive scalar, phase shifts and device selection are presented in the following Section \ref{alternating}.

\section{Alternating Optimization} \label{alternating}
\subsection{Transmit Power Allocation}
By virtue of the channel estimation methods in \cite{Wei2021Channel, Liu2020Matrix}, it is assumed that the global CSI is available to the BS, and each device has the knowledge of the receive scalar $a$ and its own CSI.
Then, with fixed phase shifts $\boldsymbol{\theta}$ in problem (\ref{MSE_minimization}), the optimal transmit power at the $k$-th device is given in a closed-form expression as follows.

In order to minimize the $\text{MSE} (\hat{s}, s)$ in (\ref{MSE_minimization_objective}), i.e., $ \sum_{k=1}^{K} |{a} \bar{h}_{k} {p_{k}} / \sqrt{\eta} - 1 | ^{2} = 0$, the optimal transmit power at the $k$-th device can be designed as
\begin{equation} \label{optimal_transmit_power}
p_{k}^{*} = \sqrt{\eta} \frac{ \left( a \bar{h}_{k} \right)  ^{\rm H} }{ \left| a \bar{h}_{k} \right| ^{2} }, \ \forall k \in \mathcal{K},
\end{equation}
where channel conditions $\bar{h}_{k}$ in (\ref{optimal_transmit_power}) can be further tuned by the multiple geo-distributed RISs, thereby reducing the energy consumption of transmitters and prolonging network lifetime.
Combining the transmit power constraints for all devices in (\ref{bi_criterion_power_constraint}), the normalizing factor $\eta$ can be calculated by
\begin{equation} \label{normalizing_factor}
\eta = P_0 \min_{k} \left| a \bar{h}_{k} \right|^{2},
\end{equation}
which is determined by the maximum transmit power of each device and the minimum equivalent channel gain $\left| a \bar{h}_{k} \right|^{2}$ of all devices.
Furthermore, taking the closed-form expression (\ref{optimal_transmit_power}) into accounts, it can be noted that each device only needs its own CSI $\bar{h}_k$, the normalizing factor $\eta$ and the receive scalar $a$ to determine the optimal transmit power $p_{k}$.
More specifically, $\bar{h}_k$ can be estimated with the downlink multicasting pilots, $\eta$ and $a$ are calculated at the BS then broadcasted to all devices.
Therefore, one of the key contributions of this work can be stated in the remark below.

\begin{remark}
	\emph{The designs for the transmit power at the device and the normalizing factor at the BS help a lot to avoid the massive signaling overhead of global CSI feedback.
	Namely, it can work in a communication-efficient manner and is beneficial to save bandwidth and alleviate congestion for the resource-limited wireless networks.}
\end{remark}

Combining (\ref{optimal_transmit_power}) and (\ref{normalizing_factor}), the MSE measurement in (\ref{MSE_s}) is further rewritten as
\begin{equation} \label{re_MSE}
\text{MSE} (\hat{s}, s) = \frac{ \sigma^2 | {a} |^2 }{ P_0 \min \limits_{k} \left| a \bar{h}_{k} \right|^{2} }.
\end{equation}

Thus, the MSE minimization problem (\ref{MSE_minimization}) can be reformulated as
\begin{eqnarray} \label{re_MSE_minimization}
&\min \limits_{ a, \boldsymbol{\theta} } & \max \limits_{k} \frac{ \sigma^2 | {a} |^2 }{ P_0 \left| a \bar{h}_{k} \right|^{2} } \quad {\rm s.t. \ (\ref{bi_criterion_phase_constraint})}.
\end{eqnarray}

\begin{proposition}\label{multi_antenna_transmit_power}
	\emph{When the BS is equipped with $N_r$ antennas, the receive vector can be denoted by $\boldsymbol{a} \in \mathbb{C}^{N_r \times 1}$ and the combined channel vector becomes $\boldsymbol{\bar h}_k \in \mathbb{C}^{N_r \times 1}$.
	Then, similar to the solutions obtained in (\ref{optimal_transmit_power}), the optimal transmit power at the $k$-th device and the normalizing factor at the BS can be derived as
	\begin{eqnarray}
	& p_{k}^{*}& = \ \sqrt{\eta} \frac{ \left( \boldsymbol{a}^{\rm H} \boldsymbol{\bar{h}}_{k} \right)  ^{\rm H} }{ \left\| \boldsymbol{a}^{\rm H} \boldsymbol{\bar{h}}_{k} \right\| ^{2} }, \ \forall k \in \mathcal{K}, \\
	& \eta& = \ P_0 \min_{k} \left\| \boldsymbol{a}^{\rm H} \boldsymbol{\bar{h}}_{k} \right\|^{2},
	\end{eqnarray}
	where $\boldsymbol{\bar h}_k = \boldsymbol{h}_k + \sum_{\ell  = 1}^L \boldsymbol{\bar G}_{\ell} \boldsymbol{\Theta}_\ell \boldsymbol{g}_k^{\ell}$,
	$\boldsymbol{h}_k \in \mathbb{C}^{N_r \times 1}$ is the channel vector between the BS and the $k$-th device,
	and $\boldsymbol{\bar G}_{\ell} \in \mathbb{C}^{N_r \times M}$ is the channel matrix from the $\ell$-th RIS to the BS.}
\end{proposition}

\subsection{Receive Scalar Control}
To facilitate analysis and derivation, we transform the min-max problem (\ref{re_MSE_minimization}) into a minimization problem with non-convex quadratic constraints.
More precisely, the problem (\ref{re_MSE_minimization}) is equivalent to the following non-linear minimization problem with non-convex quadratic constraint:
\begin{subequations} \label{MIQP}
	\begin{eqnarray}
	\label{minimize_a}
	&\min \limits_{ a, \boldsymbol{\theta} }&  | {a} |^2 \\
	\label{equivalent_channel_gain_constraint}
	&{\rm s.t.}& \left| a \bar{h}_{k} \right|^{2} \ge 1, \forall k \in \mathcal{K}, \\ 
	&{}& {\rm (\ref{bi_criterion_phase_constraint})}.
	\end{eqnarray}
\end{subequations}
Then, we provide the closed-form solutions for the optimal receive scalar and the optimal reflection matrix in the following theorem.

\begin{theorem} \label{receive_scalar_and_reflection_matrix}
	\emph{The optimal receive scalar $a^{*}$ to problem (\ref{MIQP}) can be given by
	\begin{equation} \label{optimal_receive_scalar}
	\left| a^{*} \right| = \frac{1}{\min \limits_{k} \left| \bar{h}_{k} \right|}.
	\end{equation}
	Meanwhile, the optimal reflection matrix $\mathbf{\Theta}_{\ell}^{*}$ satisfies
	\begin{equation} \label{implicit_reflection}
	\arg \left( \sum \nolimits_{\ell=1}^{L} \mathbf{\bar{g}}_{\ell} \mathbf{\Theta}_{\ell}^{*} \mathbf{g}_{k}^{\ell} \right) = \arg \left( h_{k} \right), \forall k \in \mathcal{K},
	\end{equation}
	where $\arg(\cdot)$ is a function that returns the phase shift of the input complex number.}
\end{theorem}

\begin{IEEEproof}
	See Appendix \ref{proof_of_theorem_1}.
\end{IEEEproof}

It can be noticed that the objective value of problem (\ref{MIQP}) just depends on the amplitude of the receive scalar $a$, we thus only need to optimize $|a|$ and the phase shift of $a$ can be arbitrary, which is confirmed by the closed-form solution (\ref{optimal_receive_scalar}) obtained in \textbf{Theorem \ref{receive_scalar_and_reflection_matrix}}.
Furthermore, due to the implicit expression in (\ref{implicit_reflection}), one can know that the optimal reflection matrix $\mathbf{\Theta}_{\ell}^{*}$ is not unique, the approach to find a feasible one will be proposed in Section \ref{subsection_phase_shifts_design}.
As an extension, the receiving control problem in the multi-antenna case at the BS is given below.

\begin{corollary} \label{corollary_mutli_antenna_MIQP}
	\emph{Considering the multi-antenna case with the solutions derived in \textit{\textbf{Proposition \ref{multi_antenna_transmit_power}}}, the problem (\ref{MIQP}) can be rewritten as}
	 	\begin{subequations} \label{mutli_antenna_MIQP}
	 	\begin{eqnarray}
	 	\label{mutli_antenna_minimize_a}
	 	&\min \limits_{ \boldsymbol{a}, \boldsymbol{\theta} }&  \left\| {\boldsymbol{a}} \right\|^2 \\
	 	\label{mutli_antenna_equivalent_channel_gain_constraint}
	 	&{\rm s.t.}& \left\| \boldsymbol{a}^{\rm H} \boldsymbol{\bar{h}}_{k} \right\|^{2} \ge 1, \forall k \in \mathcal{K}, \\ 
	 	&{}& (\ref{bi_criterion_phase_constraint}).
	 	\end{eqnarray}
	 \end{subequations}
\end{corollary}

According to the problem (\ref{mutli_antenna_MIQP}) in \textbf{Corollary \ref{corollary_mutli_antenna_MIQP}}, when the phase shifts $\boldsymbol{\theta}$ is fixed, the subproblem of receive vector control in the multi-antenna case can be written as
\begin{equation} \label{mutli_antenna_QCQP}
\underset{\boldsymbol{a}}{\rm \min} \ \left\| {\boldsymbol{a}} \right\|^2 \quad {\rm s.t. \ (\ref{mutli_antenna_equivalent_channel_gain_constraint})}.
\end{equation}
%However, the non-convex constraint (\ref{mutli_antenna_equivalent_channel_gain_constraint}) makes it intractable to solve the quadratically constrained quadratic programming (QCQP) problem (\ref{mutli_antenna_QCQP}) straightforwardly.
To address the non-convexity of problem (\ref{mutli_antenna_QCQP}), an intuitive approach is to reformulate it as a semidefinite programming (SDP) problem using the matrix lifting technique.
Specifically, we first define $\mathbf{H}_{k} = \boldsymbol{\bar h}_{k} \boldsymbol{\bar h}_{k}^{\rm H}$ and $\mathbf{A} = \boldsymbol{a} \boldsymbol{a}^{\rm H}$, while satisfying $\mathbf{A} \succeq \boldsymbol{0}$ and ${\rm rank}(\mathbf{A}) = 1$.
Thereby, problem (\ref{mutli_antenna_QCQP}) can be equivalently reformulated as the following matrix optimization problem with a rank-one constraint:
\begin{subequations} \label{mutli_antenna_matrix_optimization}
	\begin{eqnarray}
	& \min \limits_{ \mathbf{A} } & {\rm tr} \left( \mathbf{A} \right) \\
	\label{mutli_antenna_trace_AH}
	& {\rm s.t.} & {\rm tr} \left( \mathbf{A} \mathbf{H}_{k} \right) \ge 1, \forall k \in \mathcal{K}, \\
	\label{mutli_antenna_matrix_A}
	& {} & \mathbf{A} \succeq \boldsymbol{0}, \\
	\label{mutli_antenna_rank_one_constraint}
	& {} & {\rm rank}(\mathbf{A}) = 1.
	\end{eqnarray}
\end{subequations}

By applying the SDR technique to simply drop the non-convex rank-one constraint (\ref{mutli_antenna_rank_one_constraint}), the problem (\ref{mutli_antenna_matrix_optimization}) can be rewritten as
\begin{subequations} \label{mutli_antenna_SDR}
	\begin{eqnarray}
	&\underset{ \mathbf{A} }{ \min }& {\rm tr} \left( \mathbf{A} \right) \\
	&{\rm s.t.}& {\rm (\ref{mutli_antenna_trace_AH}) \ {\rm and} \ (\ref{mutli_antenna_matrix_A})},
	\end{eqnarray}
\end{subequations}
which is convex and can be efficiently solved by existing optimization solvers such as CVX \cite{Grant2014CVX}.
If the obtained optimal solution $\mathbf{A}^{*}$ satisfies ${\rm rank}(\mathbf{A}^{*}) = 1$, the corresponding optimal receive scaling vector $\boldsymbol{a}^{*}$ can be recovered by $\mathbf{A}^{*} = \boldsymbol{a}^{*} \boldsymbol{a}^{\rm *H} $.
Whereas, it is worth noting that if ${\rm rank}(\mathbf{A}^{*}) \ne 1$, a near-optimal rank-one solution can be calculated by $\tilde{\mathbf{A}}^{*} = \lambda \mathbf{u} \mathbf{u}^{\rm H}$ to approximate the optimal higher-rank solution $\mathbf{A}^{*}$, where $\mathbf{u}$ is the eigenvector of $\mathbf{A}^{*}$, and $\lambda$ is the corresponding maximum eigenvalue.
In the sequel, the suboptimal receive scaling vector $\tilde{\boldsymbol{a}}^{*}$ can be approximately obtained as $\tilde{\boldsymbol{a}}^{*} = \sqrt{\lambda} \mathbf{u}$.
Alternatively, the Gaussian randomization method \cite{Luo2010Semidefinite} can be adopted as a surrogate approach to similarly obtain a feasible solution to problem (\ref{mutli_antenna_matrix_optimization}), if the higher-rank solution $\mathbf{A}^{*}$ obtained by solving (\ref{mutli_antenna_SDR}) fails to be rank-one.

To overcome the limitations brought by dropping the rank-one constraint directly, and with the aim of alleviating the performance loss when the SDR is not tight for problem (\ref{mutli_antenna_matrix_optimization}), we instead propose the SCA method to solve problem (\ref{mutli_antenna_QCQP}).
First of all, we introduce the following auxiliary variables to represent the real part and the imaginary part of $\boldsymbol{a}^{\rm H} \boldsymbol{\bar{h}}_{k}$ as
\begin{equation} \label{mutli_antenna_variable_b}
\mathbf{b}_{k} = [\bar{x}_{k}, \ \bar{y}_{k}]^{\rm T}, \ \forall k \in \mathcal{K},
\end{equation}
where $\bar{x}_{k} = {\rm Re} ( \boldsymbol{a}^{\rm H} \boldsymbol{\bar{h}}_{k} )$, $\bar{y}_{k} = {\rm Im} ( \boldsymbol{a}^{\rm H} \boldsymbol{\bar{h}}_{k} )$.
Then, the non-convex constraint (\ref{mutli_antenna_equivalent_channel_gain_constraint}) becomes $\left\| \boldsymbol{a}^{\rm H} \boldsymbol{\bar{h}}_{k} \right\|^{2} = \left\| \mathbf{b}_{k} \right\|^{2} \ge 1, \ \forall k \in \mathcal{K}$, which is still non-convex.
When problem (\ref{mutli_antenna_MIQP}) is solved by the iterative approach, we use the first-order Taylor expansion to approximate the lower bound of the non-convex part based on the previous iteration, which is given by
\begin{equation} \label{mutli_antenna_Taylor_constraint}
\left\| \mathbf{b}_{k} \right\|^{2} \ge \left\| \mathbf{b}_{k}^{(z)}  \right\|^{2}
+ 2 \left( \mathbf{b}_{k}^{(z)} \right)^{\rm T} \left( \mathbf{b}_{k} - \mathbf{b}_{k}^{(z)} \right) \ge 1, \ \forall k \in \mathcal{K},
\end{equation}
where $\mathbf{b}_{k}^{(z)}$ is one feasible solution at the $z$-th iteration.

Consequently, using the auxiliary variables (\ref{mutli_antenna_variable_b}) and replacing (\ref{mutli_antenna_equivalent_channel_gain_constraint}) with its approximation (\ref{mutli_antenna_Taylor_constraint}) during each iteration, the non-convex problem (\ref{mutli_antenna_QCQP}) can be approximated by
\begin{equation} \label{mutli_antenna_SCA}
\underset{ \boldsymbol{a}, \{\mathbf{b}_{k}\} }{\rm \min} \ \left\| {\boldsymbol{a}} \right\|^2 \quad {\rm s.t. \ (\ref{mutli_antenna_variable_b}) \ and \ (\ref{mutli_antenna_Taylor_constraint})},
\end{equation}
which is convex and can be efficiently solved by CVX as well.
Note that the initial solutions of $\boldsymbol{a}^{(0)}$ and $\{ \mathbf{b}_{k}^{(0)} \}$ are found by solving the SDP problem (\ref{mutli_antenna_SDR}).
Then, the performance is continuously enhanced by resolving problem (\ref{mutli_antenna_SCA}) in an iterative fashion.
Thus, based on the above analysis for the multi-antenna case at the BS, the SDR-based algorithm for receive vector control can be summarized in \textbf{Algorithm \ref{mutli_antenna_SDR_for_receive_scaling}}.

\begin{algorithm}[t]
	\caption{SDR-Based Algorithm for Receiving Control}
	\label{mutli_antenna_SDR_for_receive_scaling}
	\begin{algorithmic}[1]
		\renewcommand{\algorithmicrequire}{\textbf{Initialize}}
		\renewcommand{\algorithmicensure}{\textbf{Output}}
		\STATE \textbf{Initialize} the tolerance $\epsilon$, maximum iteration number $N_{1}$, and the current iteration $n_1=0$.
		\STATE Given $\boldsymbol{p}$ and $\boldsymbol{\theta}$, compute $\mathbf{A}^{*}$ by solving (\ref{mutli_antenna_SDR}).
		\IF{${\rm rank}(\mathbf{A}^{*}) = 1$}
		\STATE Recover $\boldsymbol{a}^{*}$ by rank-one decomposition $\mathbf{A}^{*} = \boldsymbol{a}^{*} \boldsymbol{a}^{\rm *H} $.
		\ELSE
		\STATE Calculate the eigen-decomposition $\tilde{\mathbf{A}}^{*} = \lambda \mathbf{u} \mathbf{u}^{\rm H}$.
		\STATE Obtain $\boldsymbol{a}^{(0)} = \tilde{\boldsymbol{a}}^{*} = \sqrt{\lambda} \mathbf{u}$.
		\STATE Derive $\{ \mathbf{b}_{k}^{(0)} \} = \left\{ \left[ {\rm Re} (\tilde{\boldsymbol{a}}^{\rm *H} \boldsymbol{\bar{h}}_{k}), {\rm Im} (\tilde{\boldsymbol{a}}^{\rm *H} \boldsymbol{\bar{h}}_{k}) \right] | \forall k \right\}$.
		\REPEAT
		\STATE Compute $\boldsymbol{a}^{(n_1 + 1)}$ and $\{ \mathbf{b}_{k}^{(n_1 + 1)} \}$ by solving (\ref{mutli_antenna_SCA}).
		\STATE Update $n_{1} := n_{1} + 1$.
		\UNTIL $|\boldsymbol{a}^{(n_1)}-\boldsymbol{a}^{(n_1 - 1)}|^{2} < \epsilon$ or $n_{1} > N_{1}$.
		\ENDIF
		\STATE \textbf{Output} the optimal $\boldsymbol{a}^{*}$ or the converged solution $\boldsymbol{a}^{(n_1)}$. 
	\end{algorithmic}
\end{algorithm}

\subsection{Phase Shifts Design} \label{subsection_phase_shifts_design}
Although the implicit expression of the optimal reflection matrix has been given in (\ref{implicit_reflection}), it is still difficult to search an optimal solution due to its non-uniqueness and the curse of dimensionality.
Therefore, it is necessary to develop an efficient method to solve the problem of phase shifts design suboptimally.
Specifically, given the receive scalar $a$, the problem (\ref{MIQP}) is reduced to a feasibility-check problem and can be reformulated as
\begin{subequations} \label{single_antenna_feasibility_check}
	\begin{eqnarray}
	&\underset{ \boldsymbol{\theta} }{\rm find}& \boldsymbol{\theta} \\
	&{\rm s.t.}& {\rm (\ref{bi_criterion_phase_constraint}) \ and \ (\ref{equivalent_channel_gain_constraint}) }.
	\end{eqnarray}
\end{subequations}

Since only a feasible solution of the phase shifts $\boldsymbol{\theta}$ can be obtained by solving problem (28), it remains unknown whether the objective value of (18) will monotonically decrease or not over iterations.
From the closed-form solution obtained in (19), one can know that the value of $\min _{k} \left| \bar{h}_{k} \right|$ should be maximized to enforce the reduction of the receive scalar $|a|$ over iterations.
To this end,  we transform the above feasibility-check problem (\ref{single_antenna_feasibility_check}) into a max-min problem with an explicit objective to enforce the reduction of $|a|$ for achieving better performance and faster convergence.
As a result, the problem (\ref{single_antenna_feasibility_check}) is rewritten as
\begin{eqnarray} \label{single_antenna_max_min}
	\underset{ \boldsymbol{\theta} }{\rm max} \ \min _{k \in \mathcal{K}} \ \left| \bar{h}_{k} \right| \quad {\rm s.t. \ (\ref{bi_criterion_phase_constraint}) }.
\end{eqnarray}

Then, we introduce an auxiliary variable $\beta  =  \min _{k \in \mathcal{K}} \ \left| \bar{h}_{k} \right|$ to further transform the max-min problem (\ref{single_antenna_max_min}) into a joint maximization problem w.r.t. $\boldsymbol{\theta}$ and $\beta $, which is given by
\begin{subequations} \label{single_antenna_max_min_transformed}
	\begin{eqnarray}
	&\underset{ \boldsymbol{\theta}, \beta  }{\rm max}& \beta  \\
	\label{max_min_equivalent_channel_gain_constraint}
	&{\rm s.t.}& \left| \bar{h}_{k} \right|^{2} \ge \beta , \forall k \in \mathcal{K}, \\
	&{}& {\rm (\ref{bi_criterion_phase_constraint})}.
	\end{eqnarray}
\end{subequations}

It is obvious that both the objective and constraints are linear functions for $\beta $, but the quadratically constraint (\ref{max_min_equivalent_channel_gain_constraint}) is non-convex for $\boldsymbol{\theta}$. Additionally, due to the uncertainty of phase rotation \cite{Wu2019IRS},
the problem (\ref{single_antenna_max_min_transformed}) cannot be straightforwardly transformed into a tractable second-order cone programming (SOCP) optimization problem.
Therefore, we combine the penalty method and SCA technique to approximately solve it in the following content.

Let $v_{\ell}^{m} = e^{j \theta_{\ell}^{m}}$, then the equivalent channel fading after receiver scaling w.r.t. the $\ell$-th RIS for the $k$-th device can be denoted as
\begin{equation} \label{variable_changes}
\mathbf{\bar{g}}_{\ell} \mathbf{\Theta}_{\ell} \mathbf{g}_{k}^{\ell} = \boldsymbol{\Phi}_{k}^{\ell} \mathbf{v}_{\ell},
\end{equation}
where $\mathbf{v}_{\ell} = [ e^{j \theta_{\ell}^{1}}, e^{j \theta_{\ell}^{2}}, \ldots, e^{j \theta_{\ell}^{M}} ]^{\rm T}$
and $\boldsymbol{\Phi}_{k}^{\ell} = \mathbf{\bar{g}}_{\ell} {\rm diag} \left( \mathbf{g}_{k}^{\ell} \right)$.

As such, the constraint (\ref{max_min_equivalent_channel_gain_constraint}) is transformed as
\begin{equation} \label{max_min_substitution}
\begin{aligned}
\left| h_{k} + \sum \nolimits_{\ell=1}^{L} \mathbf{\bar{g}}_{\ell} \mathbf{\Theta}_{\ell} \mathbf{g}_{k}^{\ell} \right|^{2} =
\left| h_{k} + \sum \nolimits_{\ell=1}^{L} \boldsymbol{\Phi}_{k}^{\ell} \mathbf{v}_{\ell} \right|^{2} \ge \beta , \ \forall k.
\end{aligned}
\end{equation}

With the above substitutions (\ref{max_min_substitution}), the joint maximization problem (\ref{single_antenna_max_min_transformed}) can be rewritten as
\begin{subequations} \label{max_min_phase_shifts_design}
	\begin{eqnarray}
	&\underset{ \mathbf{v}, \beta  }{\rm max}& \beta  \\
	\label{max_min_phase_shift_constraint}
	&{\rm s.t.}& | v_{\ell}^{m} | = 1, \ \forall \ell, m, \\
	 \label{max_min_equivalent_channel_fading_constraint}
	&{}& \left| h_{k} + \boldsymbol{\Phi}_{k} \mathbf{v} \right|^{2} \ge \beta , \ \forall k,
	\end{eqnarray}
\end{subequations}
where $\mathbf{v} = [\mathbf{v}_{1}, \mathbf{v}_{2}, \ldots, \mathbf{v}_{L}] ^{\rm H}$ and
$\boldsymbol{\Phi}_{k} = [ \boldsymbol{\Phi}_{k}^{1}, \boldsymbol{\Phi}_{k}^{2}, \ldots, \boldsymbol{\Phi}_{k}^{L} ]$.
Although the constraints and variables in (\ref{max_min_phase_shifts_design}) are changed, and different from those in (\ref{single_antenna_max_min_transformed}), it is still difficult to obtain the global optimum solution due to the non-convex constraints (\ref{max_min_phase_shift_constraint}) and (\ref{max_min_equivalent_channel_fading_constraint}).

To handle the non-convexity of constraint (\ref{max_min_phase_shift_constraint}), we use the penalty function method to reformulate the problem (\ref{max_min_phase_shifts_design}) as follows:
\begin{subequations} \label{maximize_penalty}
	\begin{eqnarray} \label{max_objective_with_penalty}
	&\underset{ \mathbf{v}, \beta  }{\rm max}& \beta  + \zeta \sum_{\ell=1}^{L} \sum_{m=1}^{M} \left( | v_{\ell}^{m} |^{2} - 1 \right) \\
	\label{phase_shift_constraint_with_penalty}
	&{\rm s.t.}& | v_{\ell}^{m} | \le 1, \ \forall \ell, m, \\
	\label{equivalent_channel_fading_constraint_with_penalty}
	&{}& \rm (\ref{max_min_equivalent_channel_fading_constraint}),
	\end{eqnarray}
\end{subequations}
where $\zeta > 0$ is a positive penalty parameter.
Note that an optimal solution to problem (\ref{maximize_penalty}) can be obtained when the punished component $\left( | v_{\ell}^{m} |^{2} - 1 \right)$ in the objective function (\ref{max_objective_with_penalty}) is enforced to be zero.
Otherwise, it can be claimed that the obtained solution can be further improved over iterations.

By applying the SCA method to deal with the non-convex problem (\ref{maximize_penalty}), the objective function (\ref{max_objective_with_penalty}) is approximated by
$\beta  + 2 \zeta \sum_{\ell=1}^{L} \sum_{m=1}^{M} {\rm Re} ( ( v_{\ell}^{m (z)} ) ^{\rm H} ( v_{\ell}^{m} - v_{\ell}^{m (z)} ) )$
where $v_{\ell}^{m (z)}$ is the obtained value of variable $\mathbf{v}$ at the $z$-th iteration.
Meanwhile, the non-convex constraint (\ref{max_min_equivalent_channel_fading_constraint}) can be replaced with its first-order Taylor approximations as follows: 
\begin{align} \label{approximated_constraint_with_penalty}
\left| h_{k} + \boldsymbol{\Phi}_{k} \mathbf{v} \right|^{2} & \ge 2 {\rm Re} \left( \left( h_{k} + \boldsymbol{\Phi}_{k} \mathbf{v}^{(z)} \right) ^{\rm H} \boldsymbol{\Phi}_{k} \left( \mathbf{v} - \mathbf{v}^{(z)} \right) \right) \nonumber \\
& + \left| h_{k} + \boldsymbol{\Phi}_{k} \mathbf{v}^{(z)} \right|^{2} \ge \beta , \ \forall k \in \mathcal{K},
\end{align}
where $\mathbf{v}^{(z)}$ is the converged value at the $z$-th iteration.

Therefore, when we replace (\ref{max_objective_with_penalty}) and (\ref{max_min_equivalent_channel_fading_constraint}) with their approximations, the problem (\ref{maximize_penalty}) can be approximated by the following one:
\begin{subequations} \label{convex_max_penalty}
	\begin{align}
	\underset{ \mathbf{v}, \beta  }{\rm \max }& \quad \beta  + 2 \zeta \sum_{\ell=1}^{L} \sum_{m=1}^{M} {\rm Re} \left( ( v_{\ell}^{m (z)} ) ^{\rm H} ( v_{\ell}^{m} - v_{\ell}^{m (z)} ) \right) \\
	{\rm s.t.}& \quad {\rm (\ref{phase_shift_constraint_with_penalty}) \ and \ (\ref{approximated_constraint_with_penalty})}.
	\end{align}
\end{subequations}
Since the objective function (36a) is linear and the feasible set with constraint (36b) is convex, problem (36) is a jointly convex optimization problem w.r.t. variables $\mathbf{v}$ and $\beta$.
The details of using the SCA method to solve problem (\ref{convex_max_penalty}) at each iteration are summarized in \textbf{Algorithm \ref{SCA_for_phase_shifts}}.
Analogous to the previous analysis, the developed SCA-based algorithm can be extended to the multi-antenna case without much effort, thus the details are omitted here for brevity.

\begin{algorithm}[t]
	\caption{SCA-Based Algorithm for Phase Shifts Design}
	\label{SCA_for_phase_shifts}
	\begin{algorithmic}[1]
		\renewcommand{\algorithmicrequire}{\textbf{Initialize}}
		\renewcommand{\algorithmicensure}{\textbf{Output}}
		\STATE \textbf{Initialize} $\mathbf{v}^{(0)}$, $\beta ^{(0)}$, the tolerances $\epsilon_1$ and $\epsilon_2$, the maximum iteration number $N_{2}$, and set the current iteration number as $n_2=1$.
		\REPEAT
		\STATE Compute $(\mathbf{v}^{(n_2)}, \beta ^{(n_2)})$ by solving problem (\ref{convex_max_penalty}).
		\STATE Calculate $\delta_1 = 2 \zeta \sum_{\ell} \sum_{m} {\rm Re} ( ( v_{\ell}^{m (n_{2} - 1)} ) ^{\rm H} ( v_{\ell}^{m (n_{2})} - v_{\ell}^{m (n_{2} - 1)} ) )$.
		\STATE Calculate $\delta_2 = \beta ^{(n_2)} - \beta ^{(n_2 - 1)}$.
		\STATE Update $n_{2} := n_{2} + 1$.
		\UNTIL ($\delta_1^{2} \le \epsilon_1$ and $\delta_2^{2} \le \epsilon_2$) or $n_{2} > N_{2}$.
		\STATE \textbf{Output} the converged solutions $\mathbf{v}^{(n_2)}$ and $\beta ^{(n_2)}$. 
	\end{algorithmic}
\end{algorithm}

\subsection{Device Selection}
Substituting (\ref{re_MSE}) into (\ref{device_selection}), the combinatorial optimization problem w.r.t. device selection can be rewritten as
\begin{subequations}\label{re_device_selection}
	\begin{eqnarray}
	\label{re_device_selection_objective}
	&\min \limits_{ \mathcal{K} } &  \frac{ \sigma^2 | {a} |^2 }{ P_0 \min \limits_{k} \left| a \bar{h}_{k} \right|^{2} } - \gamma \left| \mathcal{K} \right| \\
	\label{re_MSE_requirement}
	&{\rm s.t.}& |a|^{2} - \rho \left| a \bar{h}_{k} \right|^{2} \le 0, \ \forall k \in \mathcal{K}, \\
	&{}&1 \le \left| \mathcal{K} \right| \le N,
	\end{eqnarray}
\end{subequations}
where $\rho = \varepsilon_0 P_0 / \sigma^{2} $ is a constant.

The objective function (\ref{re_device_selection_objective}) is not only related to the set cardinality $\left| \mathcal{K} \right|$, but also depends on the minimum equivalent channel gain.
Thus, solving this minimization problem (\ref{re_device_selection}) is highly intractable as it requires a complex combinatorial optimization where the elements in $\mathcal{K}$ directly affects both the value of $\min _{k} \left| a \bar{h}_{k}\right|^{2}$ and the number of feasible constraints (\ref{re_MSE_requirement}).
To support efficient algorithm design, we propose to reformulate the problem (\ref{re_device_selection}) as a joint optimization problem presented in the following lemma.
\begin{lemma} \label{lemma_device_selection}
	\emph{Let $\tau = \frac{ \bar{\rho} }{ \min _{k} \left| a \bar{h}_{k} \right|^{2} }$, where $\bar{\rho} = \frac{ \sigma^2 | {a} |^2 }{ \gamma P_0 }$. Then, the problem (\ref{re_device_selection}) can be equivalently transformed into the following joint maximization problem:}
	\begin{subequations}\label{max_re_device_selection}
		\begin{eqnarray}
		\label{max_re_device_selection_objective}
		&\max \limits_{ \mathcal{K}, \tau } & \left| \mathcal{K} \right| - \tau \\
		\label{tau_linear}
		&{\rm s.t.}& \bar{\rho} - \tau \left| a \bar{h}_{k} \right|^{2} \le 0, \ \forall k \in \mathcal{K}, \\
		\label{max_re_MSE_requirement}
		&{}& 1 - \rho \left| \bar{h}_{k} \right|^{2} \le 0, \ \forall k \in \mathcal{K}, \\
		&{}& 1 \le \left| \mathcal{K} \right| \le N.
		\end{eqnarray}
	\end{subequations}
\end{lemma}

\begin{IEEEproof}
	See Appendix \ref{proof_of_lemma_3}.
\end{IEEEproof}

Note that a trade-off relationship between $\left| \mathcal{K} \right|$ and $\tau$ is formed in problem (\ref{max_re_device_selection}).
Specifically, if the number of feasible constraints is increased (i.e., a larger $\left| \mathcal{K} \right|$), then the value of $\tau$ in (\ref{tau_linear}) shall be larger as well, which may make the objective value decrease, and vice versa.
To solve this non-trivial problem, we first introduce an auxiliary vector $\boldsymbol{e} = [e_1, e_2, \ldots, e_N] \in \mathbb{R}_{+}^{N}$, then the problem (\ref{re_device_selection}) can be equivalently reformulated as
\begin{subequations}\label{min_non_zero}
	\begin{eqnarray}
	\label{min_non_zero_objective}
	&\min \limits_{ \boldsymbol{e} \in \mathbb{R}_{+}^{N}, \tau } & \left\| \boldsymbol{e} \right\|_{0} + \tau \\
	\label{min_tau_linear}
	&{\rm s.t.}& \bar{\rho} - \tau \left| a \bar{h}_{k} \right|^{2} \le e_{k}, \ \forall k \in \mathcal{K}, \\
	\label{min_non_zero_MSE}
	&{}& 1 - \rho \left| \bar{h}_{k} \right|^{2} \le e_{k}, \ \forall k \in \mathcal{K}, \\
	&{}&1 \le \left| \mathcal{K} \right| \le N,
	\end{eqnarray}
\end{subequations}
where $\left\| \boldsymbol{e} \right\|_{0} $ is the $\ell_{0}$ norm and is equal to the number of non-zero elements in $\boldsymbol{e}$, $\mathbb{R}_{+}^{N}$ denotes the non-negative space of $1 \times N$ real-valued vector

Thus, it can be known from (\ref{min_non_zero}) that the $n$-th device should be selected to participate in the model uploading process if $e_n =0, n=1, \ldots, N$.
To handle the non-convexity of (\ref{min_non_zero_objective}), the $\ell_{0}$ norm can be rewritten as the difference of two convex functions, which is given by \cite{Gotoh2018DC}
\begin{equation} \label{KF_norm}
\left\| \boldsymbol{e} \right\|_{0} = \min \left\lbrace k : \left\| \boldsymbol{e} \right\|_{1} - |\!|\!| \boldsymbol{e} |\!|\!|_{k} = 0, \ 0 \le k \le N \right\rbrace,
\end{equation}
where $\left\| \boldsymbol{e} \right\|_{1}$ is the $\ell_{1}$ norm and is calculated by the sum of all absolute values, 
$|\!|\!| \boldsymbol{e} |\!|\!|_{k}$ is the Ky Fan $k$ norm and is obtained by the sum of largest-$k$ absolute values.

Replacing (\ref{min_non_zero_objective}) with (\ref{KF_norm}), problem (\ref{min_non_zero}) is expressed as the DC programming problem
\begin{subequations}\label{DC}
		\begin{eqnarray}
		\label{DC_objective}
		&\min \limits_{ \boldsymbol{e}, \tau } & \left\| \boldsymbol{e} \right\|_{1}+ \tau - |\!|\!| \boldsymbol{e} |\!|\!|_{k} \\
		\label{DC_MSE}
		&{\rm s.t.}& {\rm (\ref{min_tau_linear}) \ and \ (\ref{min_non_zero_MSE})}, \\
		\label{DC_succeq}
		&{}& \boldsymbol{e} \succeq \boldsymbol{0}.
		\end{eqnarray}
\end{subequations}
where $\boldsymbol{e} \succeq \boldsymbol{0}$ denotes that all elements in vector $\boldsymbol{e}$ are greater than or equal to zero.

Although problem (\ref{DC}) is non-convex, it can be solved by the majorization-minimization algorithm \cite{Sun2017MM} in an iterative fashion.
To ensure a convergent solution, we add quadratic terms to make both $\tilde{g}$ and $\tilde{h}$ be $\alpha$-strongly convex functions.
Meanwhile, the indicator function $I(\boldsymbol{e})$ can be denoted by
\begin{equation} \label{}
I(\boldsymbol{e}) =
\left\lbrace 
\begin{array}{ll}
0, & \ \text{if} \ \boldsymbol{e} \succeq \boldsymbol{0},  \\
+ \infty, & \ \text{otherwise}.
\end{array}
\right.
\end{equation}

Then, the DC objective (\ref{DC_objective}) is rewritten as the difference of two strongly convex functions, i.e., $\tilde{g} - \tilde{h}$, which can be given by
\begin{equation}
\tilde{f} = \tilde{g} - \tilde{h} = \left\| \boldsymbol{e} \right\|_{1}+ \tau - |\!|\!| \boldsymbol{e} |\!|\!|_{k} + I(\boldsymbol{e}),
\end{equation}
where $\tilde{g} = \left\| \boldsymbol{e} \right\|_{1}+ \tau + \frac{\alpha}{2} \left\| \boldsymbol{e} \right\|_{F}^{2} + I(\boldsymbol{e})$
and $\tilde{h} = |\!|\!| \boldsymbol{e} |\!|\!|_{k} + \frac{\alpha}{2} \left\| \boldsymbol{e} \right\|_{F}^{2}$.

By replacing the non-convex part $\tilde{h}$ with its linear approximation, problem (\ref{DC}) can be reconstructed as the following jointly convex optimization problem
\begin{subequations}\label{convex_DC}
	\begin{eqnarray}
	\label{convex_DC_objective}
	&\min \limits_{ \boldsymbol{e}, \tau } & \tilde{g} - \langle \partial_{\boldsymbol{e}^{(z)}} \tilde{h}, \boldsymbol{e} \rangle \\
	\label{convex_DC_MSE}
	&{\rm s.t.}& {\rm (\ref{min_tau_linear}), \ (\ref{min_non_zero_MSE}) \ and \ (\ref{DC_succeq})},
	\end{eqnarray}
\end{subequations}
where
$\boldsymbol{e}^{(z)}$ is the converged solution at the $z$-th iteration,
$\partial_{\boldsymbol{e}^{(z)}} \tilde{h}$ is the subgradient of $\tilde{h}$ w.r.t. $\boldsymbol{e}$ at $\boldsymbol{e}^{(z)}$,
and $\langle \partial_{\boldsymbol{e}^{(z)}} \tilde{h}, \boldsymbol{e} \rangle$ denotes the inner product of two vectors.
Specifically, the subgradient of $\tilde{h}$ w.r.t. $\boldsymbol{e}$ can be given by
\begin{equation}
\partial_{\boldsymbol{e}} \tilde{h} = \partial |\!|\!| \boldsymbol{e} |\!|\!|_{k} + \alpha \boldsymbol{e},
\end{equation}
where the $n$-th entry of $\partial |\!|\!| \boldsymbol{e} |\!|\!|_{k}$ can be computed by
\begin{equation}
\partial |\!|\!| \boldsymbol{e} |\!|\!|_{k} = 
\left\lbrace 
\begin{array}{l}
{\rm sign}\left( e_n \right), \ |e_n| \ge |e_{(k)}|, \\
0, \quad \quad \quad \ \ |e_n| < |e_{(k)}|.
\end{array} 
\right .
\end{equation}
The proposed DC-based algorithm for solving problem (\ref{convex_DC}) is summarized in \textbf{Algorithm \ref{DC_algorithm}}.
Additionally, the process of using DC programming to solve the device selection problem in the multi-antenna case at the BS can be developed similarly, which is omitted here for brevity.
Till now, one can see that the decoupled subproblems have been transformed into the solvable range of existing optimization toolkits one by one.

\begin{algorithm}[t]
	\caption{DC-Based Algorithm for Device Selection}
	\label{DC_algorithm}
	\begin{algorithmic}[1]
		\renewcommand{\algorithmicrequire}{\textbf{Initialize}}
		\renewcommand{\algorithmicensure}{\textbf{Output}}
		\STATE \textbf{Initialize} $\boldsymbol{e}^{(0)}$, $\tau^{(0)}$, the tolerance $\epsilon$, the maximum iteration number $N_{3}$, and set $n_3=0$.
		\REPEAT
		\STATE Calculate the subgradient $\partial_{\boldsymbol{e}^{(n_{3})}} \tilde{h}$.
		\STATE Compute the inner product $\langle \partial_{\boldsymbol{e}^{(n_{3})}} \tilde{h}, \boldsymbol{e} \rangle$.
		\STATE Obtain ($\boldsymbol{e}^{(n_3 + 1)}, \tau^{(n_3 + 1)}$ ) by solving problem (\ref{convex_DC}).
		\STATE Update $n_{3} := n_{3} + 1$;
		\UNTIL the decrease value of (\ref{convex_DC_objective}) is below $\epsilon$ or $n_{3} > N_{3}$.
		\STATE \textbf{Output} the converged solution $(\boldsymbol{e}^{(n_3)}, \tau^{(n_3)})$.
	\end{algorithmic}
\end{algorithm}

\section{Convergence and Complexity} \label{convergence}
According to the derived closed-form solutions and the developed iterative algorithms in the previous section, an alternating optimization algorithm is proposed to solve the original challenging problem (\ref{bi_criterion}).
The overall algorithm framework for dealing with the single-antenna case is given in \textbf{Algorithm \ref{alternating_optimization}}.
In the first step, the transmit power at each device is performed based on the closed-form solution derived in (\ref{optimal_transmit_power}), and the normalizing factor is calculated by (\ref{normalizing_factor}).
In the second step, the receive scalar at the BS is controlled by the closed-form solution obtained in (\ref{optimal_receive_scalar}).
In the third step, the phase shifts at each RIS is determined according to the SCA-based reflection design algorithm, i.e., \textbf{Algorithm \ref{SCA_for_phase_shifts}}.
In the fourth step, the devices participating in the model updating process are selected by the BS based on the DC algorithm, i.e., \textbf{Algorithm \ref{DC_algorithm}}. 
Previously, a holistic flowchart of our designed alternating optimization algorithm has been given in Fig. \ref{flowchart}.
In addition, the alternating optimization algorithm for solving the problems in the multi-antenna case is analogous to the processes of \textbf{Algorithm \ref{alternating_optimization}}, the main differences are \rmnum{1}) replacing the closed-form expressions (\ref{optimal_transmit_power}) and (\ref{normalizing_factor}) with the solutions obtained in \textbf{Proposition \ref{multi_antenna_transmit_power}}; \rmnum{2}) solving the subproblem (\ref{mutli_antenna_QCQP}) to obtain $\boldsymbol{a}$ using \textbf{Algorithm \ref{mutli_antenna_SDR_for_receive_scaling}}; \rmnum{3}) extending \textbf{Algorithm \ref{SCA_for_phase_shifts}} and \textbf{Algorithm \ref{DC_algorithm}} to the multi-antenna cases.
In the following context, the convergence and complexity of our proposed four-step \textbf{Algorithm \ref{alternating_optimization}} are analyzed.

\begin{algorithm}[t]
	\caption{Alternating Optimization Algorithm for Solving Problem (\ref{bi_criterion})}
	\label{alternating_optimization}
	\begin{algorithmic}[1]
		\renewcommand{\algorithmicrequire}{\textbf{Initialize}}
		\renewcommand{\algorithmicensure}{\textbf{Output}}
		\STATE \textbf{Initialize} a feasible solution $( \boldsymbol{p}^{(0)}, a^{(0)}, \boldsymbol{v}^{(0)}, \boldsymbol{e}^{(0)} )$, the maximum iteration number is denoted by $N_{4}$, and set the current iteration number as $n_4=0$.
		\REPEAT
		\STATE \textbf{Step 1:} transmit power allocation
		\STATE Given $( a^{(n_4)}, \boldsymbol{v}^{(n_4)}, \boldsymbol{e}^{(n_4)} )$, calculate $\boldsymbol{p}^{(n_4+1)}$ and $\eta^{(n_4+1)}$ by using the derived closed-form expressions in (\ref{optimal_transmit_power}) and (\ref{normalizing_factor}).
		\STATE \textbf{Step 2:} receive scalar control
		\STATE Given $( \boldsymbol{p}^{(n_4+1)}, \boldsymbol{v}^{(n_4)}, \boldsymbol{e}^{(n_4)} )$, calculate $a^{(n_4+1)}$ by using the closed-form solution in (\ref{optimal_receive_scalar}).
		\STATE \textbf{Step 3:} phase shifts design
		\STATE Given $( \boldsymbol{p}^{(n_4+1)}, a^{(n_4+1)}, \boldsymbol{e}^{(n_4)} )$, solve the reflection design subproblem (\ref{convex_max_penalty}) to obtain $\boldsymbol{v}^{(n_4+1)}$ by using {Algorithm \ref{SCA_for_phase_shifts}}.
		\STATE \textbf{Step 4:} device selection
		\STATE Given $( \boldsymbol{p}^{(n_4+1)}, a^{(n_4+1)}, \boldsymbol{v}^{(n_4+1)} )$, solve the device selection subproblem (\ref{convex_DC}) to obtain $\boldsymbol{e}^{(n_4+1)}$ by using {Algorithm \ref{DC_algorithm}}.
		\STATE Update $n_{4} := n_{4} + 1$.
		\UNTIL the objective value of (\ref{bi_criterion}) converges or $n_{4} > N_{4}$.
		\STATE \textbf{Output} the converged solution $( \boldsymbol{p}^{(n_4)}, a^{(n_4)}, \boldsymbol{v}^{(n_4)}, \boldsymbol{e}^{(n_4)} )$. 
	\end{algorithmic}
\end{algorithm}

\subsection{Convergence}
In \textbf{Algorithm \ref{alternating_optimization}}, we denote $(  \boldsymbol{p}^{(z)}, a^{(z)}, \boldsymbol{v}^{(z)}, \boldsymbol{e}^{(z)} )$ as the solution to problem (\ref{bi_criterion}) obtained at the $z$-th iteration, where the objective value is defined as
\begin{equation} \label{objective_value}
	U^{(z)} = U \left( \boldsymbol{p}^{(z)}, a^{(z)}, \boldsymbol{v}^{(z)}, \boldsymbol{e}^{(z)} \right).
\end{equation}

Substituting $(  \boldsymbol{p}^{(z)}, a^{(z)}, \boldsymbol{v}^{(z)}, \boldsymbol{e}^{(z)} )$ into (\ref{MSE_minimization}), and executing \textbf{Step 1-2-3-4} once again, we have
{\allowdisplaybreaks[4]
\begin{subequations}
\begin{eqnarray} \label{iteration_value}
&{}&U \left( \boldsymbol{p}^{(z)}, a^{(z)}, \boldsymbol{v}^{(z)}, \boldsymbol{e}^{(z)} \right) \\
&\overset{(a)}{\ge}& U \left( \boldsymbol{p}^{(z+1)}, a^{(z)}, \boldsymbol{v}^{(z)}, \boldsymbol{e}^{(z)} \right) \\
&\overset{(b)}{\ge}& U \left( \boldsymbol{p}^{(z+1)}, a^{(z+1)}, \boldsymbol{v}^{(z)}, \boldsymbol{e}^{(z)} \right) \\
&\overset{(c)}{=}& U \left( \boldsymbol{p}^{(z+1)}, a^{(z+1)}, \boldsymbol{v}^{(z+1)}, \boldsymbol{e}^{(z)} \right) \\
&\overset{(d)}{\ge}& U \left( \boldsymbol{p}^{(z+1)}, a^{(z+1)}, \boldsymbol{v}^{(z+1)}, \boldsymbol{e}^{(z+1)} \right), 
\end{eqnarray}
\end{subequations}
where the inequality (a) comes from the fact that the transmit power $\boldsymbol{p}^{(z+1)}$ obtained in \textbf{Step 1} enforces $ |{a} \bar{h}_{k} {p_{k}} / \sqrt{\eta} - 1 | $ to be zero, which can be confirmed in (\ref{optimal_transmit_power}).}
The inequality (b) holds since $a^{(z+1)}$ is obtained by solving (\ref{MIQP}) in \textbf{Step 2}, which further minimizes the MSE value.
Afterwards, the equality (c) is satisfied by finding a feasible solution of phase shifts $\boldsymbol{v}^{(z+1)}$ in \textbf{Step 3}, it can be noticed from (\ref{MIQP}) that the value of $\boldsymbol{v}^{(z+1)}$ is not related to the objective function (\ref{minimize_a}) when the receive scalar $a^{(z+1)}$ is obtained.
But, solving problem (\ref{single_antenna_max_min_transformed}) in the third step is conducive to continuously reducing the value of $\left| a \right|$ over iterations.
Similarly, the inequality (d) is owing to the continuous refinement of the number of selected devices in \textbf{Step 4}, which makes the objective value smaller and smaller.

Therefore, combining (\ref{objective_value}) and (\ref{iteration_value}), one can observe that the objective value of problem (\ref{bi_criterion}) is monotonically non-increasing over iterations, which can be expressed as
\begin{subequations}
\begin{align} \label{non_increasing}
	U^{(z)}
	&= U \left( \boldsymbol{p}^{(z)}, a^{(z)}, \boldsymbol{\theta}^{(z)}, \boldsymbol{e}^{(z)} \right) \\
	&\ge U \left( \boldsymbol{p}^{(z+1)}, a^{(z+1)}, \boldsymbol{\theta}^{(z+1)}, \boldsymbol{e}^{(z+1)} \right) = U^{(z+1)}.
\end{align}
\end{subequations}
Finally, due to the fact that the MSE value is lower bounded by zero and the number of device is upper bounded by $N$, thus the sequence $U^{(z)}$ is lower bounded and is capable to at least converge to a locally optimal solution of the original MINLP problem (\ref{bi_criterion}), if not an optimal solution.
Namely, it can be concluded that \textbf{Algorithm \ref{alternating_optimization}} is guaranteed to converge as long as the value of $N_4$ is set large enough.

\subsection{Complexity}
When the reformulated subproblems are solved by CVX, the interior point method is considered, unless otherwise stated.
For \textbf{Algorithm \ref{alternating_optimization}}, the main complexity of solving problem (\ref{bi_criterion}) lies in tackling the reflection design subproblem (\ref{convex_max_penalty}) with \textbf{Algorithm \ref{SCA_for_phase_shifts}} (i.e., \textbf{Step 3}) as well as dealing with the device selection subproblem (\ref{convex_DC}) with \textbf{Algorithm \ref{DC_algorithm}} (i.e., \textbf{Step 4}).
When it comes to \textbf{Algorithm \ref{SCA_for_phase_shifts}} for solving the subproblem of phase shifts design, the dimension of variables to be solved is $LM+1$.
Hence, the complexity is bounded by $\mathcal{O}\left( N_2 (LM + 1)^{3} \right)$, where $N_2$ is the maximum iteration number for checking the feasibility of phase shifts.
To solve the DC programming problem (\ref{DC}), the second-order interior point method \cite{Boyd2004Convex} is adopted by \textbf{Algorithm \ref{DC_algorithm}}, and then the computational cost can be given as $\mathcal{O}( N_{3} (N+1)^{2} )$, where $N_{3}$ is the maximum iteration number before the decrease of the objective value in (\ref{convex_DC_objective}) is below the preset threshold.
As a result, the total complexity of solving the MINLP problem (\ref{MSE_minimization}) with \textbf{Algorithm \ref{alternating_optimization}} is $\mathcal{O}_{1} = \mathcal{O} \left( N_2 N_4 (LM + 1)^{3} + N_3 N_4 (N+1)^{2} \right)$, where $N_4$ is the maximum iteration number for finding the converged objective value.

Regarding the complexity of solving problems in the multi-antenna cases, it mainly depends on \textbf{Step 2-3-4}.
In the second step, the complexity of \textbf{Algorithm \ref{mutli_antenna_SDR_for_receive_scaling}} consists of two parts: the initial process of solving the SDR problem (\ref{mutli_antenna_SDR}) and the iterative process of solving the convex problem (\ref{mutli_antenna_SCA}).
Specifically, the worst-case complexity of solving the SDR problem (\ref{mutli_antenna_SDR}) during the initialization in \textbf{Algorithm \ref{mutli_antenna_SDR_for_receive_scaling}} is $\mathcal{O}\left( (N_r^2 + K)^{3.5} \right)$ \cite{Luo2010Semidefinite}, and the complexity of solving the convex problem (\ref{mutli_antenna_SCA}) at each iteration in \textbf{Algorithm \ref{mutli_antenna_SDR_for_receive_scaling}} is $\mathcal{O}\left( (N_r + 2K)^{3} \right)$.
Thus, the overall complexity of solving problem (\ref{mutli_antenna_QCQP}) with \textbf{Algorithm \ref{mutli_antenna_SDR_for_receive_scaling}} can be represented as $\mathcal{O}\left( (N_r^2 + K)^{3.5} + N_1 (N_r + 2K)^{3} \right)$, where $N_1$ is the maximum iteration number for finding the converged receive scaling factor.
Consequently, the complexity of using alternating optimization techniques to solve problems in the multi-antenna case can be given by
$\mathcal{O}_{2} = \mathcal{O} \left( N_4 (N_r^2 + K)^{3.5} + N_1 N_4 (N_r + 2K)^{3} + N_2 N_4 (LM + 1)^{3} \notag\right. \\ \left. + N_3 N_4 (N+1)^{2} \right)$. % \notag\right. \\ \left.

\begin{figure*}[t!]
	\begin{minipage}[t]{0.32 \textwidth}
		\centering
		\includegraphics[width=2.3 in]{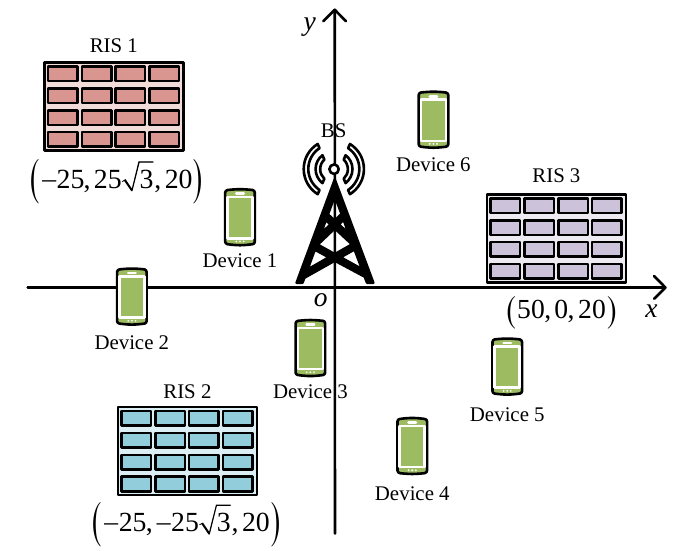}
		\caption{Simulation setup of multi-RIS aided AirFL systems (top view).}
		\label{simulation_setup}
	\end{minipage}
	\hspace{1 mm}
	\begin{minipage}[t]{0.32 \textwidth}
		\centering
		\includegraphics[width=2.3 in]{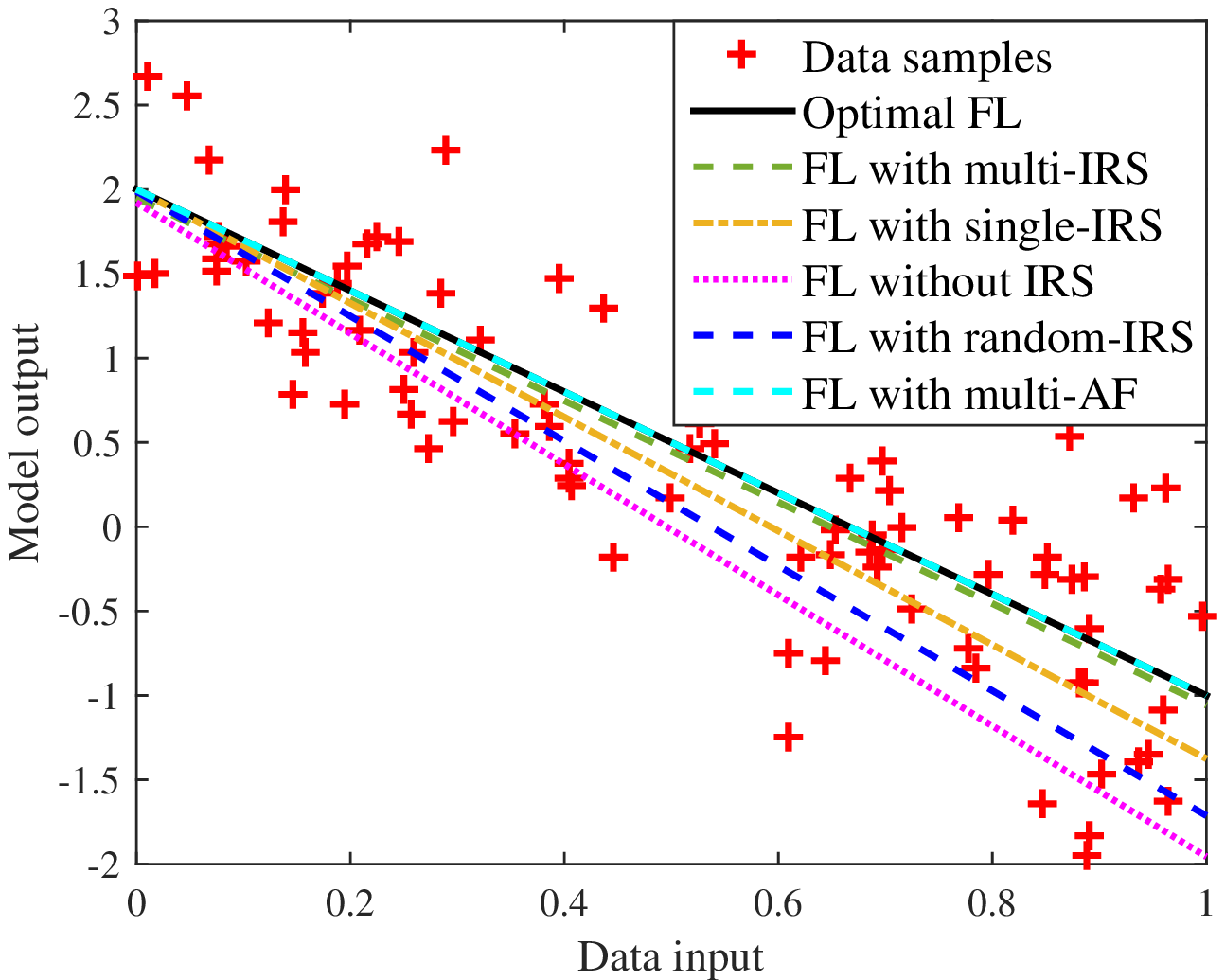}
		\caption{An example of implementing FL for linear regression.}
		\label{linear_regression}
	\end{minipage}
	\hspace{1 mm}
	\begin{minipage}[t]{0.32 \textwidth}
		\centering
		\includegraphics[width=2.3 in]{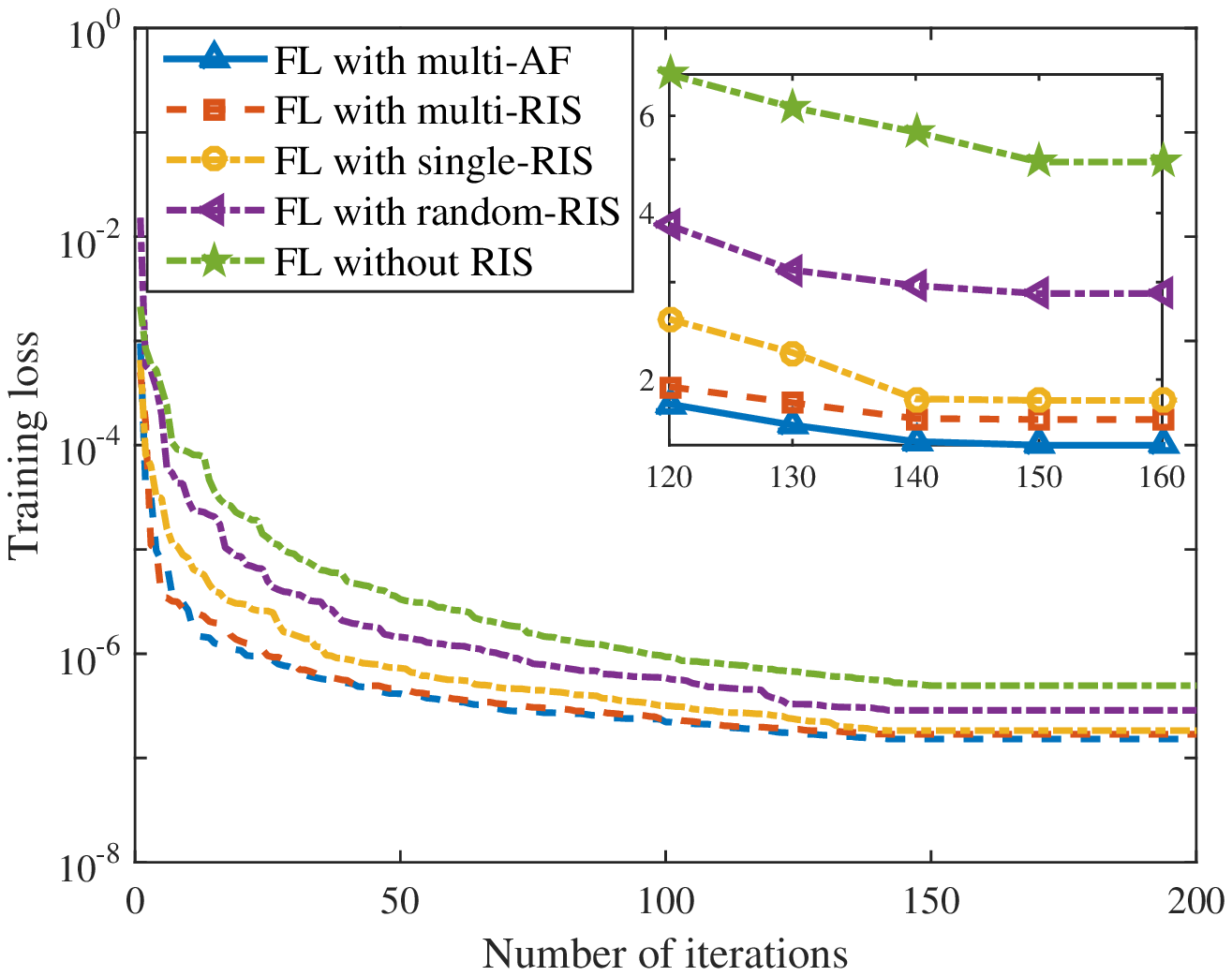}
		\caption{Training loss versus the number of iterations.}
		\label{training_loss_vs_iterations}
	\end{minipage}
\end{figure*}

\section{Experimental Settings and Results} \label{simulation}
\subsection{Simulation Settings} \label{settings}
As shown in Fig. \ref{simulation_setup}, we consider that there are $N=6$ IoT devices, $L=3$ RISs and one BS in the AirFL system, where all devices are uniformly distributed in a square area of size $100 \times 100$ (in meters) with the BS located at its center \cite{Yang2021Energy}.
In the three-dimensional (3D) Cartesian coordinates, the location of the $\ell$-th RIS is given by $(x_{\ell}, y_{\ell}, z_{\ell}) = ( 50 \cos( \frac{2 \pi \ell}{L}), 50 \sin( \frac{2 \pi \ell}{L}), 20)$, and each RIS is equipped with $M=60$ reflecting elements.
It is assumed that all devices are on the horizontal plane, and the BS is located at $(0,0,25)$.
Moreover, the maximum transmit power at each device is set as $P_{0}=23$ dBm, and the noise power is $\sigma^{2} = -80$ dBm.
Other parameters are set to $\gamma=0.2$ and $\varepsilon_0 = 0.01$.
The channel gain equals to the small-scale fading multiplied by the square root of the path loss, please refer to \cite{Ni2021Resource} for specific settings of the channel model.

In order to validate the effectiveness of our proposed algorithms for the multi-RIS aided federated leaning (labeled `FL with multi-RIS'), the FL is used to train a linear regression model to predict the relationship between $x$ and $y$ \cite{Chen2020A}.
The input $x$ and output $y$ follow the function $y = -3x + 2 + 0.5 \times n_{0}$ where the input data $x$ is randomly generated from $[0,1]$, and the Gaussian noise $n_{0}$ follows $\mathcal{N}(0,1)$.
Specifically, the regress function in the MATLAB is invoked to fit 30 on-device samples for linear regression at each iteration.
%Moreover, the proposed FL framework is also adopted to train a $7$-layered convolutional neural network (CNN) for image classification on the MNIST dataset\footnote{\url{http://yann.lecun.com/exdb/mnist/}}, and a $50$-layered residual network (ResNet) on the CIFAR-10 dataset\footnote{\url{https://www.cs.toronto.edu/~kriz/cifar.html}}.
Moreover, the proposed FL framework is also adopted to train a $7$-layered convolutional neural network (CNN) for image classification on the MNIST dataset, and a $50$-layered residual network (ResNet) on the CIFAR-10 dataset.
For comparison purposes, the following four schemes are considered as benchmarks in our experiments.
\begin{enumerate}
	\item[\rmnum{1}.] FL without RIS: There is only one BS and $N$ devices in the federated learning system, where AirComp is adopted to compute specific functions via concurrent transmission over multi-access channels.
	\item[\rmnum{2}.] FL with single-RIS: Compared with scheme 1, one central RIS is deployed at $(50,0,20)$ to assist the model uploading from devices to the BS. For the fairness of comparison, the number of reflecting elements for the central RIS equals to $L \times M$.
	\item[\rmnum{3}.] FL with random-RIS: The single RIS with random phase shifts is also considered as one benchmark.
	Note that the elements in $\boldsymbol{\theta}$ are randomly chosen from $[0, 2 \pi]$, while other variables are solved by our proposed algorithms.
	\item[\rmnum{4}.]FL with multi-AF: The deployment of multiple amplify-and-forward (AF) relays is the same as that of FL with multi-RIS scheme. Namely, there are three active AF relays that work in half-duplex mode, and each consists of $M$ antennas.
\end{enumerate}

%According to the above settings, if not specified, all numerical results are averaged over 1,000 independent Monte-Carlo simulations.

\subsection{Performance Evaluation} \label{MSE_performance}

\subsubsection{Implementing FL for linear regression}
\begin{figure*}[t!]
	\centering
	\begin{minipage}[t]{0.32 \textwidth}
		\centering
		\includegraphics[width=2.3 in]{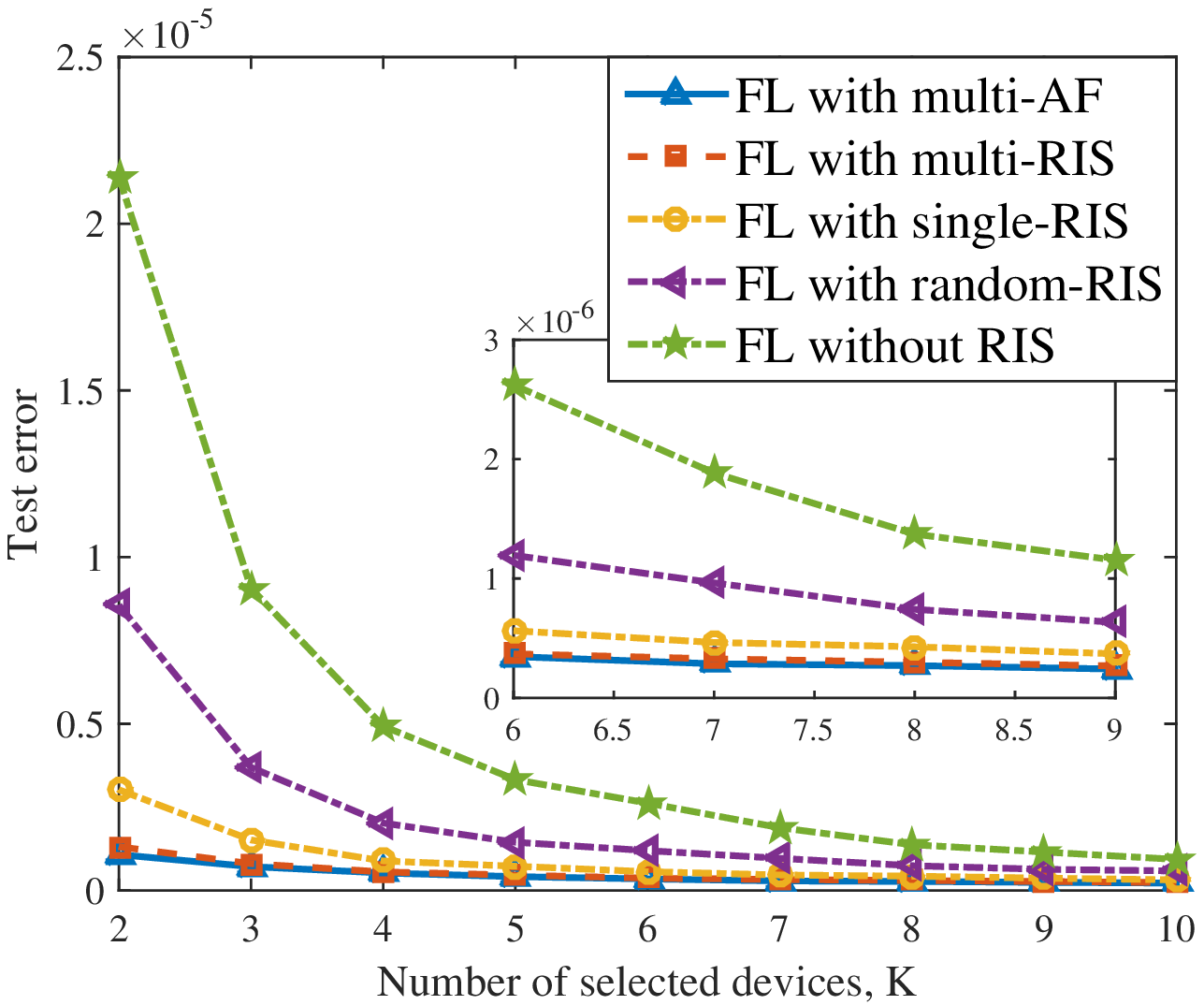}
		\caption{Test error vs. $K$.}
		\label{MSE_vs_devices}
	\end{minipage}
	\begin{minipage}[t]{0.32 \textwidth}
		\centering
		\includegraphics[width=2.3 in]{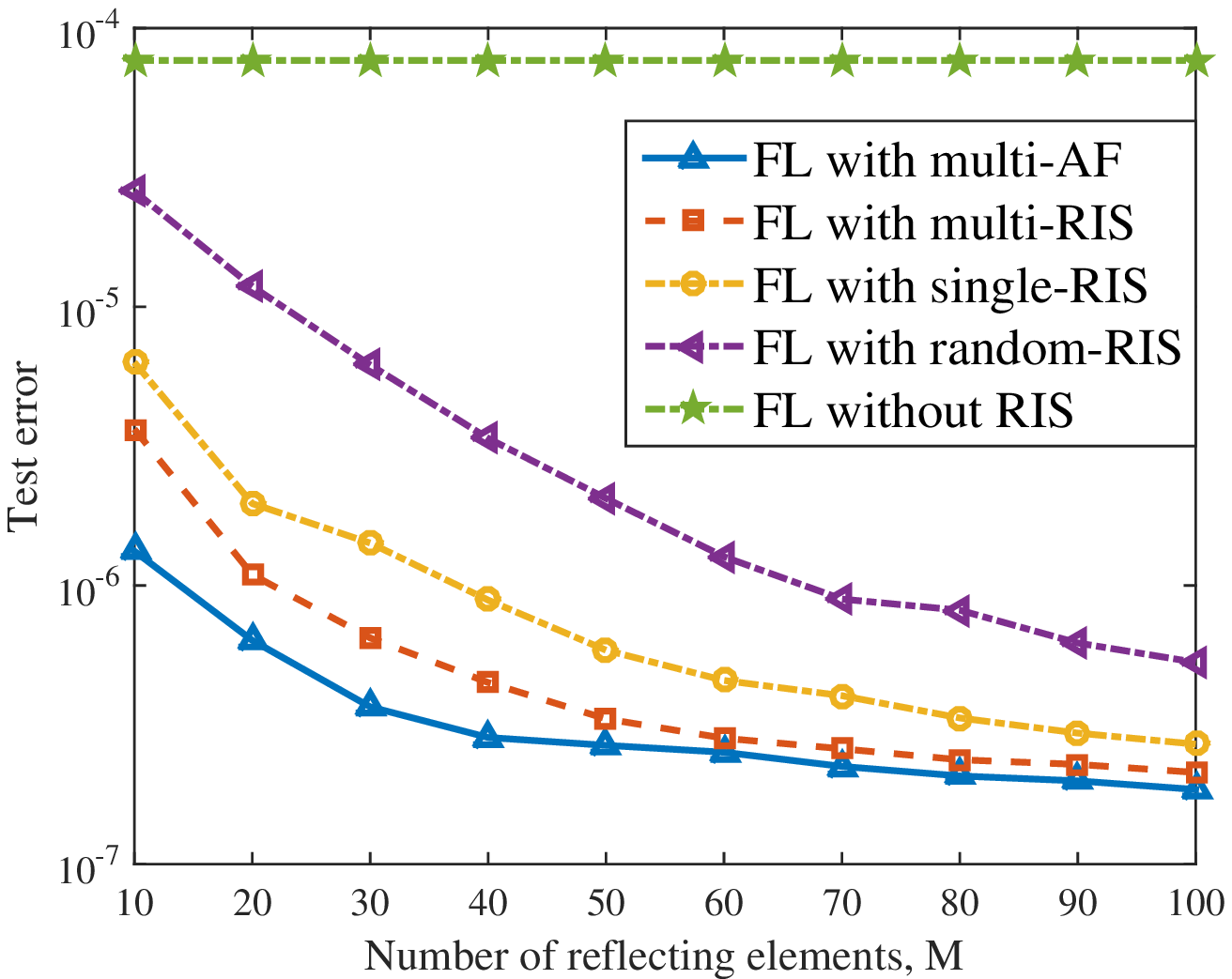}
		\caption{Test error vs. $M$.}
		\label{MSE_vs_elements}
	\end{minipage}
	\begin{minipage}[t]{0.32 \textwidth}
		\centering
		\includegraphics[width=2.3 in]{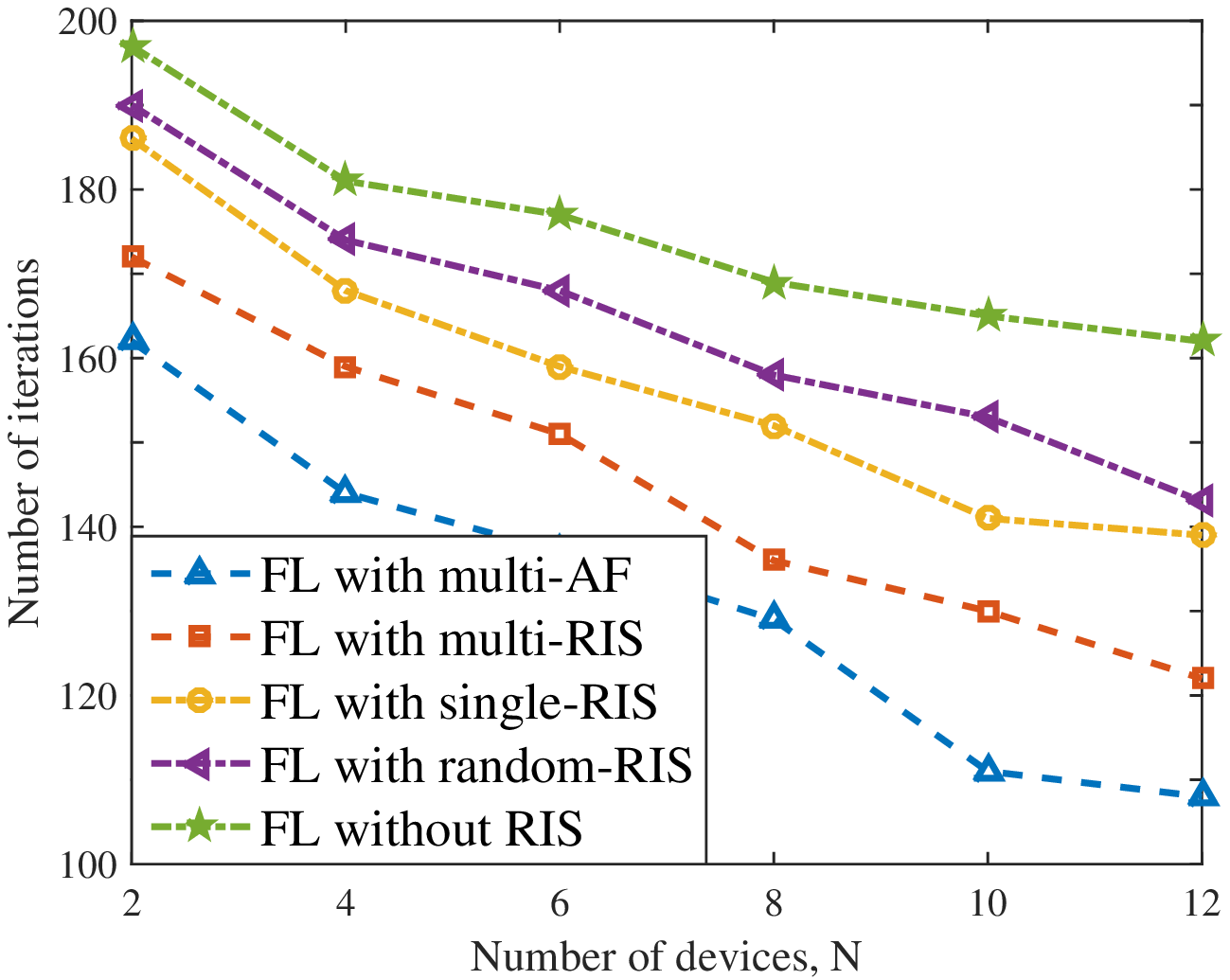}
		\caption{Number of iterations vs. $N$.}
		\label{iterations_vs_devices}
	\end{minipage}
\end{figure*}

In Fig. \ref{linear_regression}, the `optimal FL' is an ideal scheme that the communication noise between the BS and devices is zero, and the relationship between $x$ and $y$ can be perfectly modeled.
It can be observed that the proposed `FL with multi-RIS' scheme is able to train a near-optimal linear regression model close to the `optimal FL' scheme, and can fit data samples more accurately than other benchmarks (`FL with single/random-RIS' and `FL without RIS').
This is because the proposed algorithms for model aggregation not only jointly consider the learning and wireless factors, but also optimize the phase shifts of distributed multiple RISs to suppress noise.
Then, Fig. \ref{training_loss_vs_iterations} shows that the proposed scheme can converge faster to a smaller training loss, similar to the active scheme of `FL with multi-AF'.
This is due to the fact that a lower signal distortion can be achieved by judiciously reconfiguring the wireless environment with multiple distributed RISs.
In Fig. \ref{MSE_vs_devices}, it can be observed that as the number of selected devices increases, the test error of global model on the testing dataset decreases.
This comes from that the global model will become more accurate if much more data samples are learned for aggregation.
Thereby, the test error of all schemes decrease owing to the improved prediction accuracy.
One can observe from Fig. \ref{MSE_vs_elements} that the test error decreases with the number of reflecting elements (or the number of antennas of each AF).
This is due to the fact that a larger number of reflecting elements can lead to a smarter wireless environment and the propagation error induced from the channel noise can be suppressed more effectively.
At last, Fig. \ref{iterations_vs_devices} illustrates that the number of iterations decreases as the number of devices increases.
When the network size becomes larger, more devices can be selected to participate learning process, which accelerates the convergence of federated learning.
Compared with RIS-related benchmarks, the proposed scheme can work more efficiently and spend fewer communication rounds with the aided of multiple RISs.

\subsubsection{Implementing FL for image classification}
\begin{figure*}[t!]
	\centering
	\begin{minipage}[t]{0.32 \textwidth}
		\centering
		\includegraphics[width=2.3 in]{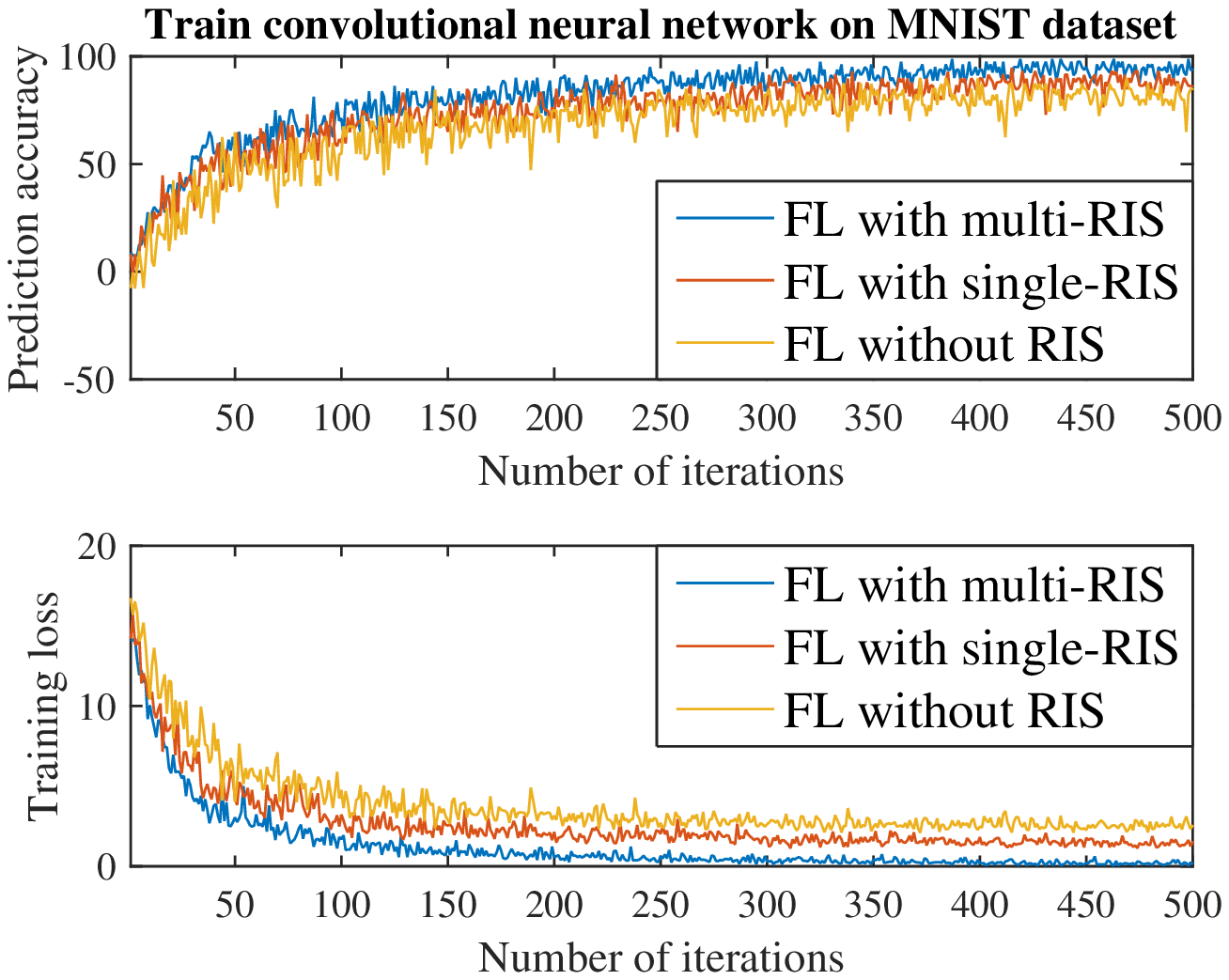}
		\caption{Train CNN on MNIST dataset.}
		\label{MNIST_loss_and_accuracy}
	\end{minipage}
	\begin{minipage}[t]{0.32 \textwidth}
		\centering
		\includegraphics[width=2.3 in]{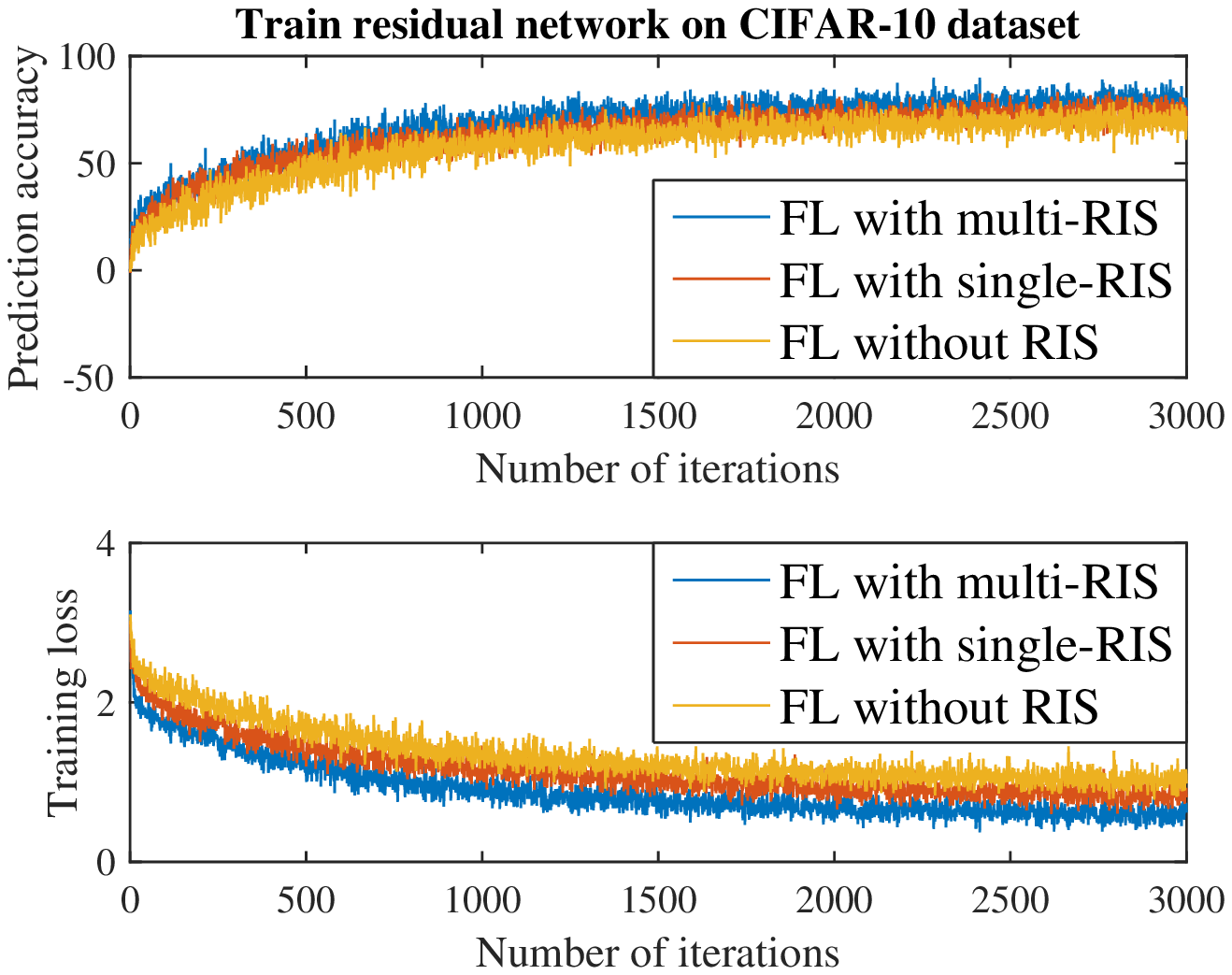}
		\caption{Train ResNet on CIFAR-10 data.}
		\label{CIFAR_loss_and_accuracy}
	\end{minipage}
	\hspace{0.1mm}
	\begin{minipage}[t]{0.32 \textwidth}
		\centering
		\includegraphics[width=2.3 in]{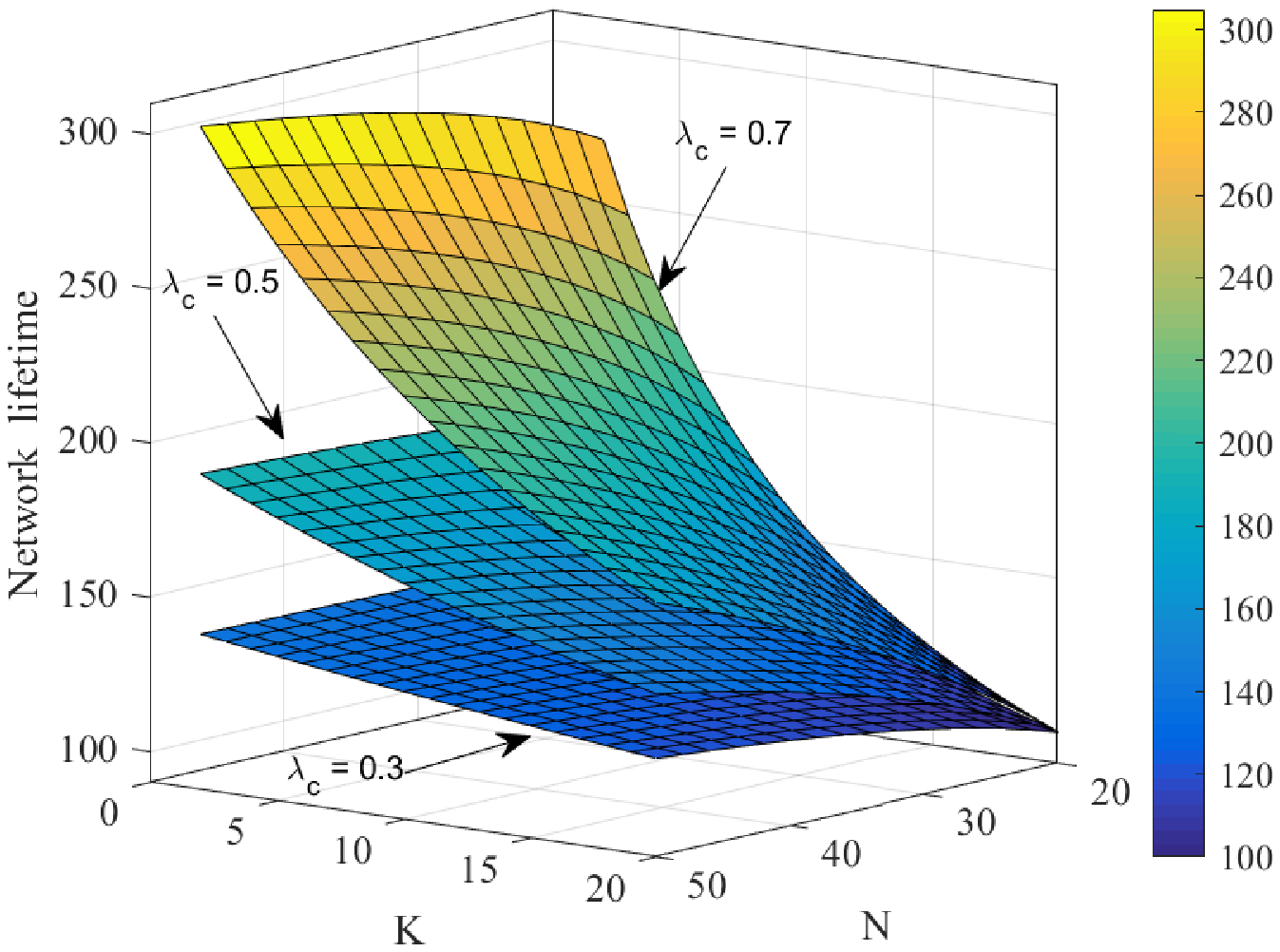}
		\caption{Network lifetime vs. $K$ and $N$.}
		\label{lifetime_vs_devices}
	\end{minipage}
\end{figure*}

In Fig. \ref{MNIST_loss_and_accuracy} and Fig. \ref{CIFAR_loss_and_accuracy}, we evaluate the learning performance for image classification on real data in terms of training loss and prediction accuracy.
Both the MNIST and CIFAR-10 datasets are divided into five training batches and one test batch, each with 10,000 images.
The on-device CNN or ResNet is trained in parallel using randomly sampled images.
To minimize loss, the stochastic gradient descent solver with an initial learning rate of 0.01 is adopted as an optimizer to update parameters at each iteration, where the size of each mini-batch is specified as 128.
Compared to benchmarks, it is noted that the proposed scheme can achieve lower training loss and higher prediction accuracy on both real datasets thanks to the reduced aggregation error with the aid of multiple RISs.

\subsubsection{Network lifetime of the considered FL system}
In Fig. \ref{lifetime_vs_devices}, the impact of various $K$ and $N$ values on the network lifetime is demonstrated.
In our simulation, if one device is selected to transmit its local parameters to the BS, it will spend $1$ unit energy for data sensing, local computing, and communication processes, during which time the percentage of total energy consumption for communications is denoted by $\lambda_{c}$.
Thus, the remaining processes require $1 - \lambda_{c}$ unit energy regardless of whether the device communicates with the BS.
Moreover, it is assumed that each device has $\delta = 100$ units energy, and the time until the first device dies is defined as the network lifetime, which can be given by $\lfloor N \delta / (N - \lambda_{c} N + \lambda_{c} K) \rfloor$, and $\lfloor \cdot \rfloor$ is the floor function.
It can be seen from this figure that a higher $\lambda_{c}$ leads to a longer network lifetime, i.e., more energy consumption for sensing and computing will shorten the network lifetime.
Additionally, one can observe that the performance of network lifetime is positively proportional to $N$ and is also inversely proportional to $K$.
Namely, if more devices are deployed and less devices are selected, a longer network lifetime can be achieved.
Therefore, the trade-off between learning performance and network lifetime is an interesting research direction in the future work.
	
\section{Conclusion} \label{conclusion}
In this paper, we investigated the model aggregation and device selection problems of federated learning in multi-RIS assisted systems by jointly optimizing the transmit power, receive scalar, phase shifts, and learning participants to minimize the aggregation error while accelerating the convergence rate of AirFL.
To solve the formulated challenging bi-criterion problem, we derived closed-form expressions for transceivers and proposed an alternating optimization algorithm to tackle the non-linear and non-convex subproblems by invoking relaxation methods such as SDR, SCA and DC programming.
Simulation results demonstrated that
\rmnum{1}) the aggregation distortion can be effectively reduced by leveraging geo-distributed intelligent surfaces to reconfigure the wireless channels,
\rmnum{2}) the learning behavior of AirFL can be improved by the designed resource allocation and device selection algorithms,
and \rmnum{3}) our alternating optimization algorithm is also capable of reducing energy consumption and prolonging network lifetime.
Despite that this paper focuses on anti-noise techniques, wireless noise is not always an obstacle and can even be regulated to enhance the generalization ability of machine learning models.
Also, the noise can be utilized to strengthen the secure communication in the parameter exchange process and reduce the privacy leakage of federated learning, which are research opportunities worthy of further exploration.

\appendices
\section{Proof of Theorem \ref{receive_scalar_and_reflection_matrix}} \label{proof_of_theorem_1}
Due to the fact that $\left| a \bar{h}_{k} \right|^{2} = \left| a \right|^{2} \left| \bar{h}_{k} \right|^{2}, \ \forall k$, the constraints (\ref{equivalent_channel_gain_constraint}) in problem (\ref{MIQP}) can be rewritten as $\left| a \right|^{2} \ge \left| \bar{h}_{k} \right|^{-2}, \ \forall k$.
Thus, the problem (\ref{MIQP}) is reformulated as
%\begin{eqnarray} \label{rewritten_receive_scalar_and_reflection_matrix}
%\min \limits_{ a, \boldsymbol{\theta} } \ | {a} |^2 \quad {\rm s.t.} \left| a \right|^{2} \ge \left| \bar{h}_{k} \right|^{-2}, \forall k \in \mathcal{K}, {\rm \ and \ (\ref{bi_criterion_phase_constraint})}.
%\end{eqnarray}
\begin{subequations} \label{rewritten_receive_scalar_and_reflection_matrix}
	\begin{eqnarray}
	&\min \limits_{ a, \boldsymbol{\theta} }& | {a} |^2 \\
	&{\rm s.t.}& \left| a \right|^{2} \ge \left| \bar{h}_{k} \right|^{-2}, \forall k \in \mathcal{K}, \\
	&{}& {\rm (\ref{bi_criterion_phase_constraint})}.
	\end{eqnarray}
\end{subequations}

It can be easily verified that at the optimal solution to problem (\ref{rewritten_receive_scalar_and_reflection_matrix}), all the constraints in (\ref{rewritten_receive_scalar_and_reflection_matrix}) should be met, i.e.,
\begin{equation} \label{theorem_optimal_receive_scalar}
	\left| a^{*} \right| = \frac{1}{\min \limits_{k} \left| \bar{h}_{k} \right|} = \max \limits_{k} {\left| {h}_{k} + \sum_{\ell=1}^{L} \mathbf{\bar{g}}_{\ell} \mathbf{\Theta}_{\ell} \mathbf{g}_{k}^{\ell} \right| ^{-1} }.
\end{equation}

Furthermore, it can be observed from (\ref{theorem_optimal_receive_scalar}) that the value of $\left| a^{*} \right|$ decreases as the value of $\left| \bar{h}_{k} \right|$ increases.
As a result, the phase shifts of RISs should be finely tuned to render the phase shift of $\sum_{\ell=1}^{L} \mathbf{\bar{g}}_{\ell} \mathbf{\Theta}_{\ell} \mathbf{g}_{k}^{\ell}$ the same as that of ${h}_{k}$ for all users, which can be expressed as
$\arg \left( \sum_{\ell=1}^{L} \mathbf{\bar{g}}_{\ell} \mathbf{\Theta}_{\ell}^{*} \mathbf{g}_{k}^{\ell} \right) = \arg \left( h_{k} \right), \forall k \in \mathcal{K}$.
This completes the proof of Theorem \ref{receive_scalar_and_reflection_matrix}.

\section{Proof of Lemma \ref{lemma_device_selection}} \label{proof_of_lemma_3}
According to the definitions of $\tau$ and $\bar{\rho}$ in Lemma \ref{lemma_device_selection}, it holds that
%$\tau = \max \limits_{k} \frac{ \bar{\rho} }{ \left| a \bar{h}_{k} \right|^{2} }$.
\begin{equation} \label{theorem_tau}
\tau = \max \limits_{k} \frac{ \bar{\rho} }{ \left| a \bar{h}_{k} \right|^{2} }.
\end{equation}
Hence, it can be observed that the value of $\tau$ should be no less than ${ \bar{\rho} }/{ \left| a \bar{h}_{k} \right|^{2} }$ for all users, i.e., $\tau \ge { \bar{\rho} } / { \left| a \bar{h}_{k} \right|^{2} }, \forall k \in \mathcal{K}$.
Then, problem (\ref{re_device_selection}) can be equivalently reformulated as
%{\allowdisplaybreaks[4]
\begin{subequations}\label{min_re_device_selection}
	\begin{eqnarray}
	&\min \limits_{ \mathcal{K}, \tau } & \tau - \left| \mathcal{K} \right| \\
	\label{constraint_tau}
	&{\rm s.t.}& \tau \ge \frac{ \bar{\rho} }{ \left| a \bar{h}_{k} \right|^{2} }, \forall k \in \mathcal{K}, \\
	&{}& |a|^{2} - \rho \left| a \bar{h}_{k} \right|^{2} \le 0, \ \forall k \in \mathcal{K}, \\
	&{}& 1 \le \left| \mathcal{K} \right| \le N,
	\end{eqnarray}
\end{subequations}
where the objective and constraints in (\ref{min_re_device_selection}) is obviously equivalent to those in (\ref{max_re_device_selection}), and thus the proof of Lemma \ref{lemma_device_selection} is completed.
Note that constraint (\ref{constraint_tau}) holds with equality for at least one $k$ of the optimal solution.

%%References are generated by ref.bib.
\bibliographystyle{IEEEtran}
\bibliography{IEEEabrv,ref}
	
% that's all folks
\end{document}